\documentclass[useAMS,usenatbib]{mnras}

\usepackage{graphicx}
\usepackage{amssymb,amsmath,multirow}
\usepackage{mathrsfs}
\usepackage{epstopdf,enumerate,url,color}
\usepackage{bm}

\title[Disc eccentricity]{Secular evolution of eccentricity in protoplanetary discs with gap-opening planets}
\author[J. Teyssandier and G.~I. Ogilvie]{Jean Teyssandier\thanks{E-mail: jt591@cam.ac.uk} and Gordon I. Ogilvie\\ 
Department of Applied Mathematics and Theoretical Physics, University of Cambridge, Cambridge CB3 0WA, United Kingdom  
}

\newcommand{\pd}[2]{\frac{\partial #1}{\partial #2}}
\newcommand{\pdd}[2]{\frac{\partial^2 #1}{\partial #2 ^2}}
\newcommand{\dd}[2]{\frac{\mathrm{d} #1}{\mathrm{d} #2}}
\newcommand{\ddd}[2]{\frac{\mathrm{d}^2 #1}{\mathrm{d} #2 ^2}}
\newcommand{\lc}[2]{b_{#1}^{(#2)}}
\newcommand{\kc}[2]{K_{#1}^{(#2)}}
\newcommand{\id}{\mathrm{d}}
\newcommand{\mi}{\mathrm{i}}
\newcommand{\me}{\mathrm{e}}

\newcommand{\rin}{r_{\mathrm{in}}}
\newcommand{\rout}{r_{\mathrm{out}}}
\newcommand{\rres}{r_{\mathrm{res}}}

\newcommand{\Mp}{M_{\rm p}}

\newcommand{\qp}{q_{\rm p}}

\newcommand{\ap}{a_{\rm p}}
\newcommand{\rp}{r_{\rm p}}
\newcommand{\tp}{T_{\rm p}}
\newcommand{\Op}{\Omega_{\rm p}}

\newcommand{\Ms}{M_*}

\newcommand{\ssc}{\Sigma_{\rm 0}}
\newcommand{\rsc}{r_{\rm 0}}

\newcommand{\cs}{c_{\rm s}}
\newcommand{\cst}{c_{\rm s}^2}
\newcommand{\Kr}{K_{\rm r}}
\newcommand{\Phip}{\Phi_{\rm p}}
\newcommand{\Ok}{\Omega_{\rm K}}

\begin{document}
\maketitle

\begin{abstract}

We explore the evolution of the eccentricity of an accretion disc perturbed by an embedded planet whose mass is sufficient to open a large gap in the disc. Various methods for representing the orbit-averaged motion of an eccentric disc are discussed. We characterize the linear instability which leads to the growth of eccentricity by means of hydrodynamical simulations. We numerically recover the known result that eccentricity growth in the disc is possible when the planet-to-star mass ratio exceeds $3\times 10^{-3}$. For mass ratios larger than this threshold, the precession rates and growth rates derived from simulations, as well as the shape of the eccentric mode, compare well with the predictions of a linear theory of eccentric discs. We study mechanisms by which the eccentricity growth eventually saturates into a non-linear regime.

\end{abstract}


\section{Introduction}
\label{sec:intro}

Planetary systems show a broad distribution of eccentricities. Studying dynamical processes taking place during disc-planet interactions is essential to understanding the mechanisms which will subsequently shape the orbital architecture of planetary systems.

Early work on the subject was conducted in the context of gap-opening satellites in planetary rings. \citet{gt80} showed that eccentric Lindblad resonances can excite the eccentricity of the satellite, while eccentric corotation resonances lead (in general) to a damping of eccentricity. The conclusion of that work was that damping caused by eccentric corotation resonances would dominate over growth by eccentric Lindblad resonances, but only by a small margin. In a following paper \citep{gt81}, these authors showed that the satellite could in turn excite the eccentricity of a disc. The consequence of the  saturation of the corotation torque was studied simultaneously by \citet{gs03} and \cite{ol03}. The results of \citet{gs03} suggested that gap-opening planets could undergo eccentricity growth if their eccentricity exceeds a small critical value.

A substantial body of numerical work exists in the context of planet-disc interactions. \citet{Papaloizou01} showed that the eccentricity of massive objects (larger than 20 Jupiter masses) could be excited by interactions with the disc, the latter also developing eccentricity. Fixing the orbit of the planet and focusing on the disc, \citet{kd06} have conducted an extensive study, and found that for typical viscosities, the disc can become significantly eccentric when the planet exceeds 3 Jupiter masses, even when the planet is held on a circular orbit. This threshold was also found by \citet{regaly10}. \citet{dlb06} have also reported eccentricity growth, both in the planet and in the disc. The growth of eccentricity of Jupiter-mass planets was also observed recently by \citet{dc15}, although the three-dimensional SPH simulations of \citet{daa13} did not show such growth, which could be the result of a small integration time. The aim of this paper is to explain and go beyond the work of \citet{kd06} by conducting a detailed analysis of the processes that lead to eccentricity excitation in a disc, while keeping the planet on a fixed circular orbit.

To this aim, in \citet{to16} we have formulated a set of linear equations that describe the propagation, excitation and damping of eccentricity in a disc-satellite system.  We have computed the precession rate and growth rate of eccentric modes in the simple case of a star orbited by a hot Jupiter in an empty cavity, with a protoplanetary disc truncated some distance outside the planet's orbit. Although the growth rate will depend on the physical parameters of the disc and the planet, we argued that eccentricity growth was possible within the disc's lifetime. In this first study, we did not consider the case of gap-opening planets, since it would require a good model for the surface density profile created by giant planets. So far, to our knowledge, all attempts to model the surface density discs in the vicinity of giant planets have failed to give the correct depth and width of the gap. In the present paper, we use hydrodynamical simulations to obtain a surface density profile, which we can use to compute eccentric modes with our linear theory. We can also directly measure the growth rate and eccentric rate from the simulations and provide a comparison with the linear theory. 

Planet-disc interactions also play an important role in observational features in protoplanetary discs. It is possible that a significant disc eccentricity leaves a observable footprint in the CO line profiles in emission, in the form of an asymmetry in the line profile. \citet{regaly10} conducted 2D hydrodynamical simulation of giant planets (several Jupiter mass) in discs, and computed the resulting observable asymmetry in the CO line profiles. A consequent asymmetry could help characterize giant planets located at a few AU in protoplanetary discs. \citet{flaherty15} reported a small asymmetry in the CO line profile of the HD 163296 protoplanetary disc, which could be associated with such an eccentric motion.

The paper is organised as follows: in Section \ref{sec:secular} we present a secular theory of eccentric discs and planets, in Section \ref{sec:num} we describe the numerical methods we use and in Section \ref{sec:orb} we discuss how to compute eccentric motion from such simulations. These three sections serve as a framework to conduct an in-depth study of eccentricity evolution, which we do in Section \ref{sec:results}. We discuss our results in Section \ref{sec:discussion}.

\section{Secular theory for the eccentricity}
\label{sec:secular}
In \citet{to16} we presented a set of linear
equations that describe the evolution of a small eccentricity during
disc-planet interactions, and applied this linear theory to
the case of a hot Jupiter in an empty cavity. In this section
we summarize some of the main results of this paper, and
detail how the model has to be modified to study the case
of a gap-opening giant planet orbiting within the disc.

\subsection{Governing equations}

The propagation and growth or decay of eccentricity in a disc are the result of various physical processes. For small eccentricities and small eccentricity gradients, neighbouring orbits  do not intersect \citep{ogilvie01}. It is possible to formulate a set of linear equations that describe the long-term (secular) evolution of the eccentricity, in a way that couples the eccentricity of the disc and that of the planet. Similarly to the secular dynamics of celestial mechanics, the equations are azimuthally-averaged, and quantities depend on the radial cylindrical polar coordinate $r$ and time $t$ only. In the simplest case, the disc is represented by a surface density $\Sigma$ which is a function of $r$ only, and a Keplerian rotation profile with angular velocity $\Omega=(GM_*/r^3)^{1/2}$. We also assume a locally isothermal disc and we define the sound speed $\cs=H\Omega$, where $H$ is the disc scale-height. We denote by $\Ms$ the mass of the star, $\Mp$ the mass of the planet, and $\qp=\Mp/\Ms$ the planet-to-star mass ratio.

Equations are formulated in terms of the complex eccentricity $E=e\, \me^{\mi \varpi}$, which in the secular theory is a function of $r$ and $t$. Below we present a list of the various physical mechanisms relevant for the linear theory:
\begin{itemize}
\item Pressure: A 2D secular linear theory of eccentric adiabatic discs was presented in \citet{go06}, and isothermal discs were discussed in \citet{to16}. In the case of a locally isothermal disc with sound speed $c_{\rm s}(r)$, the equation governing the propagation of a small eccentricity due to pressure has the form of a dispersive wave equation:
\begin{align}
\label{eq:2diso}
\Sigma r^2 \Omega \left( \pd{E}{t}\right)_{\rm pressure}  &= \frac{\mi}{2r}\pd{}{r}\left(\Sigma \cst r^3 \pd{E}{r} \right) + \frac{\mi r}{2}\dd{}{r}\left(\Sigma\cst \right)E\nonumber\\
& - \frac{\mi}{2r}\pd{}{r}\left(\Sigma \dd{\cst}{r} r^3 E\right).
\end{align}
\item Secular gravitational effect: the secular potential of a planet on a circular orbit is represented by that of a ring whose mass is that of the planet:
\begin{align}
\label{eq:epd1}
\Sigma r^2 \Omega & \left( \pd{E}{t}\right)_{\rm pd} = \mi G\Mp\Sigma(r)K_{3/2}^{(1)}(r,\ap)E,
\end{align}
where $K_{3/2}^{(1)}$ is equivalent to a Laplace coefficient and is given in Eq. (\ref{eq:ksm}). If the disc is represented as a collection of eccentric rings, then this formulation is equivalent to the classical Laplace-Lagrange theory of planetary dynamics \citep{md99}.
\item Viscosity: We follow \citet{go06} and adopt a simplified model of eccentricity damping with a Shakura-Sunyaev $\alpha$-parametrization:
\begin{equation}
\Sigma r^2 \Omega \left( \pd{E}{t} \right)_{\rm visc} = \frac{1}{2r}\pd{}{r}\left(\alpha \Sigma \cst r^3 \pd{E}{r} \right),
\end{equation}
where $\alpha$ is a dimensionless parameter. As stressed in \citet{to16}, this effective bulk viscosity accounts for damping by any thermal or mechanical process, apart from resonances which are described below.
\item Eccentric Lindblad resonances (ELR): They correspond to locations in the disc where the perturbing frequency in the rotating frame matches the epicyclic frequency (see Section \ref{sec:resloc}). They lead to a local growth of eccentricity:
\begin{align}
\label{eq:elrd}
\Sigma r^2 \Omega & \left(\pd{E}{t}\right)_{\rm ELR}  = \frac{G\Mp^2}{M_*}\Sigma\mathscr{A}^2 E w_{\rm L}^{-1}\Delta\left( \frac{r-\rres}{w_{\rm L}} \pm 1 \right).
\end{align}
Here $\Delta(x)=(2\pi)^{-1/2}\exp(-x^2/2)$ is a Gaussian representing the broadening of the resonant effect by pressure, shifted away from the nominal resonant radius $r=\rres$ by one resonance width $w_{\rm L}$ (see Section \ref{sec:reswidth}). In addition $\mathscr{A}$ is a coefficient whose expression can be found in \citet{to16} and is a function of $r$.
\item Eccentric corotation resonances (ECR): They correspond to locations in the disc where the perturbing frequency in the rotating frame is zero (see Section \ref{sec:resloc}). They can lead to either a growth or decay of eccentricity, depending on the local vortensity gradient, but the net effect is in general a damping of eccentricity. They read: 
\begin{align}
\label{eq:ecrd}
\Sigma r^2 \Omega & \left( \pd{E}{t} \right)_{\rm ECR} = \pm\dd{\ln (\Sigma/\Omega)}{\ln r}\frac{G\Mp^2}{M_*}\Sigma\mathscr{C}^2 E\nonumber\\
&\times w_{\rm C}^{-1}\Delta\left( \frac{r-\rres}{w_{\rm C}} \right).
\end{align}
Again, the broadening of ECRs over a width $w_{\rm C}$ is represented by a Gaussian function  $\Delta$ (see Section \ref{sec:reswidth}), and $\mathscr{C}$ is a coefficient whose expression can be found in \citet{to16} and is a function of $r$.
\item Boundary terms: In the hydrodynamical simulations we present later on in this paper, the velocity of fluid elements is relaxed towards a circular state at both edges of the disc, using an exponential damping (see Eq. \ref{eq:bc}). This leads to a damping of eccentricity of the form: 
\begin{equation}
\Sigma r^2 \Omega  \left( \pd{E}{t}\right)_{\rm BC} = -\frac{\Sigma r^2 \Omega}{\tau_{\rm i,o}} R_{\rm i,o}(r)E.
\label{eq:ebc}
\end{equation}
Here $\tau_{\rm i,o}$ represents the damping time at the inner and outer edge of the disc, and $R_{\rm i,o}$ is a ramp function representing the radial zone over which this boundary condition is effective (see Section \ref{sec:bc}).
\end{itemize}
In \citet{to16} we studied other mechanisms at play. The most important ones were a 3D term in the pressure equation, which we do not include here since we conduct 2D hydrodynamical simulations, and the coupling with a non-zero planet eccentricity. We also exclude the self-gravity of the disc as well as short-range forces from the present analysis. 

In order to solve equations (\ref{eq:2diso}) to (\ref{eq:ebc}), we seek normal modes of the form $E(r) \me^{\mi \omega t}$. The precession rate of the mode is given by $\Re(\omega)$, while its growth rate is $-\Im(\omega)$. The method to solve these equations is described in detail in \citet[][see also Appendix \ref{app:discretize}]{to16}. In appendix \ref{app:int}, we also give useful expressions for how the various physical processes listed above contribute to the growth rate or precession rate.

\subsection{Departure from Keplerian rotation}
\label{sec:depart}

In the gap and in the vicinity of the gap edges, strong density gradients and perturbations from the planet are likely to cause departure from Keplerian orbits. From the radial component of the momentum equation, the angular frequency is given by
\begin{equation}
\label{eq:momr}
-r\Omega^2= - \frac{1}{\Sigma}\pd{(\Sigma\cst)}{r} - \pd{\Phi}{r} ,
\end{equation}
where $\Phi$ is the gravitational potential. When neglecting the disc's self-gravity, we have $\Phi = \Phi_*$ + $\Phip$, representing the sum of the stellar and planetary potentials, respectively. The stellar potential is simply $\Phi_*=-GM_*/r$, while details on the computation of $\Phip$ and its derivatives are given in Appendix \ref{app:phip}. Here we merely state that $\Phip$ is an orbit-averaged quantity, and therefore depends only on $r$. 

We assume a locally isothermal disc with the sound speed given by $\cs=H\Ok$, with $H/r$ the constant disc aspect ratio and $\Ok=(GM_*/r^3)^{1/2}$ the Keplerian frequency at radius $r$. Equation (\ref{eq:momr}) can be rewritten so that the angular frequency is given as a function of radius by:
\begin{equation}
\label{eq:omdisc}
\Omega(r)^2 = \Ok^2\left[1-\left(\frac{H}{r}\right)^2 \right] + \frac{\cst}{r}\pd{\ln \Sigma}{r} - \frac{1}{r}\pd{\Phip}{r}.
\end{equation}
From now on, this is the frequency that we will use in Equations (\ref{eq:2diso}) to (\ref{eq:ebc}).

Departure from Keplerian orbits will also affect the epicyclic frequency $\kappa$ given by $\kappa^2 = 4\Omega^2+r\mathrm{d}\Omega^2/\mathrm{d}r$. Using eq. (\ref{eq:omdisc}) we find
\begin{align}
\label{eq:kappadisc}
\kappa(r)^2 &= \Ok^2\left[1-\left(\frac{H}{r}\right)^2\left(1+\pd{\ln \Sigma}{\ln r} \right) \right] \nonumber\\
&+ \frac{3\cst}{r}\pd{\ln \Sigma}{r} + \cst\pdd{\ln\Sigma}{r} \nonumber\\
&- \frac{3}{r}\pd{\Phip}{r} - \pdd{\Phip}{r}.
\end{align}

\begin{figure}
    \begin{center}
    \includegraphics[scale=0.9]{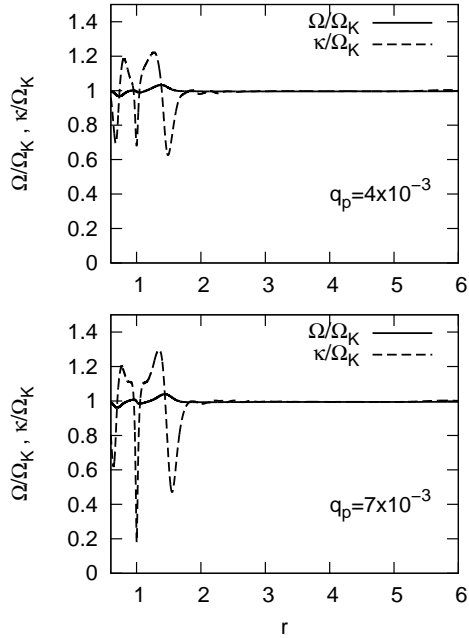}
    \caption{Angular frequency $\Omega$ (solid line) and epicyclic frequency $\kappa$ (dashed line) as a function of radius, divided by the Keplerian frequency $\Omega_{\rm K}$, for two different planet-to-star mass ratios: $\qp=4\times10^{-3}$ (top panel) and $\qp=7\times10^{-3}$ (bottom panel).}
    \label{fig:kappaom}
    \end{center}
\end{figure}

On Figure \ref{fig:kappaom} we show the departure from Keplerian orbits, in the form of $\Omega/\Omega_{\rm K}$ and $\kappa/\Omega_{\rm K}$ as a function of radius, for two different mass ratios. The parameters are the same as the ones we use in our simulations described in Section \ref{sec:num}, and the surface density is extracted from the same simulations, and can bee seen for instance in Figure \ref{fig:qp_sigma}. Departure from Keplerian orbits is significant in the disc, with a strong feature at the location of the planet due to its gravitational effect. Note that the divergence at $r=\ap$ is avoided by applying a smoothing length, representing a vertical averaging of the planet's potential (see appendix \ref{app:phip}). In addition to this strong feature, there is overall a slightly larger departure from Keplerian orbits for higher mass planets, and beyond $r=2$ the orbits are very much Keplerian. Most of the departure from Keplerian orbits take place in the gap, where the gradients of surface density and gravitational potential of the planet are strong.

\subsection{Resonance location}
\label{sec:resloc}
\begin{figure}
    \begin{center}
    \includegraphics[scale=0.65]{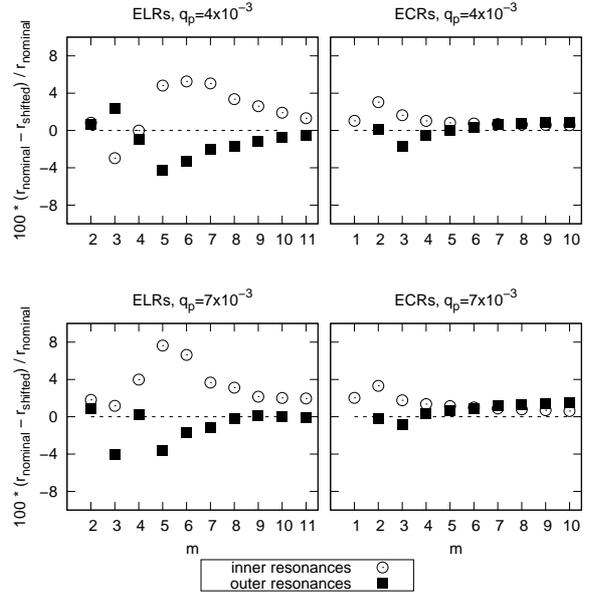}
    \caption{Shift of resonance locations for two different planet masses, $\qp=4\times10^{-3}$ (top panel), and $\qp=7\times10^{-3}$ (bottom panel). The left column is for ELRs and the right column is for ECRs. Circles represent inner resonances $(r<\ap)$ and squares represent outer resonances $(r<\ap)$. The nominal resonant radius $r_{\rm nominal}$ is the one that would be defined by setting $\Omega=\kappa=\Omega_{\rm K}$, the Keplerian frequency (i.e. the resonances occupy the radii of orbital mean motion resonances). The shifted radius $r_{\rm shifted}$ is the one computed in Section \ref{sec:resloc}. The $y$-axis shows $100(r_{\rm nominal}-r_{\rm shifted})/r_{\rm nominal}$. For outer resonances, a positive (resp., negative) value of this quantity indicates that resonances is shifted away from (resp., closer to) the planet. The effect of resonance shifting is more significant for low-$m$ outer ELRs.} 
    \label{fig:resloc}
    \end{center}
\end{figure}

For the disc interior to the planet, the location of (inner) ELRs is given by $(m+1)\Op - m\Omega(r)=-\kappa(r)$, while for the disc exterior to the planet, the location of (outer) ELRs is given by  $(m-1)\Op - m\Omega(r)=\kappa(r)$. Similarly, inner ECRs are located at $(m+1)\Op - m\Omega(r)=0$ and outer ECRs are located at $(m-1)\Op - m\Omega(r)=0$. In the case where $\Omega=\kappa=\Omega_{\rm K}$, the resonance locations reduce to those of orbital mean motion resonances. In the general case where $\Omega$ and $\kappa$ are different from $\Omega_{\rm K}$ the locations of all these resonances will be affected. For a given surface density profile, the shifted locations can be computed numerically from  equations (\ref{eq:omdisc}) and (\ref{eq:kappadisc}). We show this departure in Figure \ref{fig:resloc}. Outer Lindblad resonances can by shifted away from their nominal radius by as much as 5\%. Given the steep surface density gradients in the disc, this can bring them to locations where their effect will be strengthened or weakened, depending on whether they are shifted closer or further from the planet.

For convenience, in the remainder of the paper we will still refer to resonances as if  they were occupying the site of a mean motion resonance. For instance, we will refer to the $m=2$ outer eccentric Lindblad resonance as a 1:3 ELR.

\subsection{Resonance width}
\label{sec:reswidth}
An estimate of the width of ELRs can be derived from the dispersion relation of waves in a non-self-gravitating gas disc. In \citet{to16} we estimated it to be:
\begin{equation}
\label{eq:rwelr}
\frac{w_{\rm L}}{r}\bigg|_{\rm ELR} \approx \left( \frac{(H/r)^2}{3(m\mp 1)} \right)^{1/3}.
\end{equation}

The width of ECRs is more problematic. \citet{ol03} showed that three lengthscales are competing to set the width of the corotation resonances. The first one derives from the width of the libration zone, and depends on the amplitude of the forcing potential $\Psi$. The associated width is $w_{\rm lib}\sim\Psi^{1/2}/\Omega$. \citet{mo04} used the following prescription for the corresponding relative resonant width:
\begin{equation}
\label{eq:rwecrl}
\frac{w_{\rm lib}}{r}\bigg|_{\rm ECR} \approx  4.1\left( C_m^{\pm}me\qp \right )^{1/2},
\end{equation}
where the $C_m^{\pm}$ coefficients are of order unity and can be found in \citet{ol03}. Such expression for the resonant width is inconvenient in a linear theory as it introduces an explicit dependence on the amplitude $e$ of the eccentricity.

The second relevant lengthscale, noted $w_{\rm visc}$, is set by the viscous diffusion process across the corotation region. It reads $w_{\rm visc}=(\nu/(-m\id\Omega/\id r))^{1/3}$, where $\nu=\alpha\cs H$ is the kinematic viscosity. For a Keplerian rotation profile, it can be approximated by: 
\begin{equation}
\label{eq:rwecrv}
\frac{w_{\rm visc}}{r}\bigg|_{\rm ECR} \approx \left(\alpha \frac{(H/r)^2}{m} \right)^{1/3}.
\end{equation}
This expression conveniently compares with eq. (\ref{eq:rwelr}). 

A third lengthscale relevant to the corotation resonance is
$c_\mathrm{s}/\kappa\approx H$, which is the decay length of the
evanescent density wave generated in this region. Although this sets
the scale of the distribution of torque on the disc from the
companion, the feedback of the evanescent density wave on the eccentric
mode occurs on the shorter lengthscale(s) mentioned above. It is not
known accurately how the corotation torque should be determined in
cases where the surface density varies strongly over a distance of
order $H$ from the resonance. For unsaturated corotation resonances, we take Eq. (\ref{eq:rwecrv}) to be the relevant lengthscale on which angular momentum is transferred via the resonant interactions.

\citet{mvs87} showed the torque density of ELRs is given by an Airy function, whose peak is shifted outward from the resonant radius by about one resonant width. In \citet{to16} we have approximated this effect by assuming that the contribution of a single ELR will spread radially following a Gaussian function whose centre is offset away from the planet by one $w_{\rm L}$, and with a Gaussian width also given by $w_{\rm L}$. This is equivalent to saying that all the contributions of the Airy function cancel each others, apart from the first peak which we approximated by the aforementioned Gaussian. On the other hand, the width of ECRs we give in equation (\ref{eq:rwecrv}) is the full width of the resonance. We also assume that the effect of ECRs radially spreads in the disc following a Gaussian centred on the resonant radius, and we set the width of this Gaussian to be $w_{\rm C}=w_{\rm visc}/5$. This factor 5 is derived assuming that 99\% of the area covered by the Gaussian lies between $-w_{\rm visc}/2$ and $+w_{\rm visc}/2$. In practice, this assumption has some important consequences, as it causes the ECRs to operate over a radial width which is about 20 times narrower than that of ELRs, which in turns strongly limits the amount of damping they can provide to compete against the growth generated by ELRs.

\section{Numerical methods}
\label{sec:num}

\subsection{General remarks}
\label{sec:numgen}

Numerical simulations were conducted using the {\sc Pluto} code
\citep{mignone12}, on a two-dimensional cylindrical grid.
The resolution is $768\times 1422$ in radius and azimuth respectively, and the
radius spacing is logarithmic. This resolution ensures a constant cell aspect ratio over the grid, and nearly square cells. We use the \textit{hllc} solver with a linear reconstruction method and a second-order Runge-Kutta time-integration scheme. We have conducted various numerical tests which are detailed in Appendix \ref{app:num}.

The simulations are locally isothermal and we use a constant aspect ratio $H/r = 0.05$ throughout the disc. We assume that angular
momentum is transported by a turbulent process prescribed
by an $\alpha$-disc model with $\alpha=4\times 10^{-3}$ and thus a
kinematic viscosity $\nu=\alpha\cs H$ is applied to the disc, where
$\cs =H\Omega_{\rm K}$ is the local sound speed ($\Omega_{\rm K}$ being the 
Keplerian frequency at radius $r$).
Units are chosen such that $\Ms+\Mp=1$, the
gravitational constant is 1, and the planet is held fixed on a circular orbit at $\rp=1$ with an orbital period of $\tp=2\pi$ and orbital frequency $\Op=1$. In these units, we set the inner edge of the disc to be located
at $\rin=0.2$ and the outer edge at $\rout=6$. 

The surface density is taken to be $\Sigma=\ssc (r/\rsc)^{-1/2}$ where
$\ssc=1$ in code units. The scaling with $\ssc$ is arbitrary since self-gravity and forces acting on the planet are not considered. A floor density is applied everywhere on the grid so that the density contrast cannot go below $\Sigma_{\rm min}/\ssc=10^{-9}$.

The initial radial velocity is zero, and the initial azimuthal velocity takes into account the small departure from Keplerian orbits due to pressure using Eq. (\ref{eq:momr}) with the planet potential being zero.

Indeed we allow the mass of the planet to grow from 0 to $\Mp$ in the
first ten orbits of the simulation.
The gravitational potential $\Phi_{\rm p}$ exerted by the planet on
the disc is smoothed by a parameter $\epsilon$, such that
\begin{equation}
\Phi_{\rm p} =G\Mp \left(-\frac{1}{\sqrt{r_{\rm rel}^2+\epsilon^2}} + \frac{\bm{r}\cdot\bm{\rp}}{\rp^3} \right)
\end{equation}
where $r_{\rm rel}$ is the relative distance between the planet and the center of the grid cell, and we take $\epsilon = 0.6H$. The second term in this equation is the indirect term arising from the fact that the coordinate origin is centred on the star, and not at the centre of mass. Accretion of mass onto the planet is not considered.

\subsection{Boundary conditions} 
\label{sec:bc}

At the inner and outer edges of the disc, we follow the prescription of \citet{dvb06}, we relax the density and both velocity components towards a given value:
\begin{equation}
\label{eq:bc}
\dd{X}{t}=-\frac{X-X_0}{t_{\rm damp}}R(r),
\end{equation}
where $X$ is the surface density or both components of velocity, $X_0$ the value towards which they are relaxed, $t_{\rm damp}$ is a
damping timescale, which we take to be a hundredth of the orbital period
at the outer radius for the outer boundary, and to be the orbital period at the inner edge for the inner boundary, and $R$ is a quadratic function that increases from 0 at the
chosen damping radius to 1 at the edge. The surface density is relaxed towards its initial value. The relaxation zone at the inner edge extends from $\rin$ to $\rin+0.1$. At the outer edge, it extends from $\rout-1$ to $\rout$.

At the inner edge, both components of the velocity are relaxed towards circular orbits around the star, while at the outer edge, both components of the velocity are relaxed towards circular orbits in the centre of mass of the star-planet system. Therefore, in the grid frame, centred on the star, both the radial and azimuthal components of the velocity take non-zero values which need to be computed at each time-step in the relaxation zone where equation (\ref{eq:bc}) applies. The non-zero value of the radial velocity at the outer edge will result in a small inflow/outflow of material, and an inflow/outflow boundary condition has to be applied \citep[see][]{nelson00}.

Periodic boundary conditions are applied in the azimuthal direction.

\section{Orbital elements of a disc}
\label{sec:orb}

In this section we introduce various ways of representing Keplerian orbits in a disc, and point out a few caveats that can lead to erroneous results when not properly taken into account. These subtleties, which are often overlooked in the literature, will be useful when analysing numerical simulations.

For each cell of the grid, one can derive a set of osculating orbital elements that correspond to the instantaneous position and velocity of the cell at a given time, using the coordinates and components of the velocity of the cell. 

Perhaps the simplest way to characterize the eccentricity of the disc would be to assign an eccentricity to each cell of the grid. Then, at a given radius $r$, one can perform an azimuthal averaging of the eccentricity of all the cells at this radius. The result would be an eccentricity profile as a function of $r$. However, defining an eccentricity as a function of radius is somewhat spurious, since the radius varies along the path of a Keplerian elliptical orbit.

A more accurate way of characterizing the eccentricity profile would be to label the orbits by their semimajor axis $a$ instead of their radius. Such representation of the disc seems more natural, as it describes the disc as a set of Keplerian ellipses. Another quantity of interest is the semilactus rectum $\lambda=a(1-e^2)$. For small eccentricities, $\lambda$ will be equivalent to $a$.

Finally, we remark that orbits with the same eccentricity at the same semi-major axis could in principle have different orientations. This can be naturally taken into account by considering the eccentricity vector instead of the eccentricity itself. We define the components of the eccentricity vector by
\begin{equation}
k=e\cos\varpi \qquad h=e\sin\varpi.
\end{equation}
These are also the real and imaginary parts of the complex eccentricity $E=e {\rm e}^{\mi \varpi}$. 

Therefore we define the eccentricity of a ring (labelled by either $r$, $a$ or $\lambda$) as:
\begin{equation}
\label{eq:edisc}
e_{\rm ring} = \sqrt{\left\langle k \right\rangle^2 + \left\langle h \right\rangle^2} .
\end{equation}
Here, $\left\langle k \right\rangle$ and $\left\langle h \right\rangle$ are the values of the components of the eccentricity vector, averaged over a ring. Here $r$ is the radial coordinate of the grid, and should not be interpreted in the celestial mechanics sense of an azimuth-dependent radius along an elliptical orbit.

In anticipation of the results of our hydrodynamical simulations, we note that when the planet is released in the disc, it exerts a strong tidal field on the latter. Before the disc eventually relaxes and adjusts to the presence of the planet, some fluid elements will follow orbits that are not exactly Keplerian. These would appear to have a high eccentricity, and different orientations. By considering the eccentricity vector instead of the eccentricity, one naturally weights the eccentricity by its orientation, and we will see that the averaging process in equation (\ref{eq:edisc}) leads to a significantly smaller (and more realistic) eccentricity than what would be measured otherwise.

Despite all these precautions, representing the disc as a set of Keplerian ellipses can remain a challenge. One issue is that, in the vicinity of the planet, fluids elements are unlikely to follow Keplerian orbits around the star. The second main issue is that near the star, fluid elements are likely to follow circular orbits around the latter, while near the outer edge of the disc, they are likely to follow circular orbits around the center of mass of the star-planet system (we neglect the self-gravity of the disc).

In order to overcome these two issues, we adopt the following conventions:
\begin{itemize}
\item When computing the orbital elements of the cells of the disc, we discard all cells within the Hill radius of the planet, $r_{\rm H} = \ap(\qp/3)^{1/3}$.
\item The eccentricity of a cell is chosen to be the minimum of its eccentricity computed around the star and around the centre of mass. That way, cells close to the star follow a circular motion around it, and cells far away from the star-planet center of mass follow circular orbits around it.
\end{itemize}

\begin{figure}
    \begin{center}
    \includegraphics[scale=1]{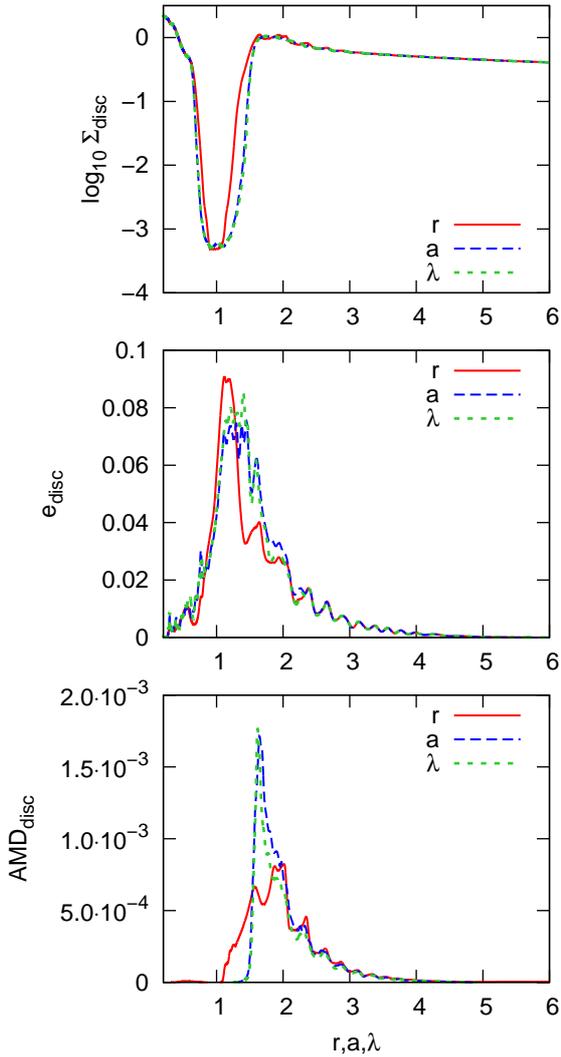}
    \caption{Surface density (top), eccentricity (middle) and AMD (bottom) for $\qp=7\times10^{-3}$, as a function of radius $r$ (solid red), semi-major axis $a$ (dashed blue) and semi-lactus rectum $\lambda$ (dotted green).}
    \label{fig:rsmaslr}
    \end{center}
\end{figure}

The procedure to derive orbit averaged properties of the disc is as follows:
\begin{itemize}
\item Each cell of the grid has two sets of osculating orbital elements, one with respect to the star, and one with respect to the centre of mass (hereafter CoM).
\item For each cell of the grid, we decide whether it is best fitted by a Keplerian orbit around the star, one around the CoM, or none of the above (e.g., the material in the Hill sphere of the planet). In the latter case, those cells are simply discarded from the analysis. Each cell now has one set of osculating elements. 
\item As mentioned already, there are at least three distances that can be used in eccentric discs: the distance $r$, the semi-major axis $a$ and the semi-lactus rectum $\lambda$. These three quantities can be measured either in a reference frame centred on the star, or on the CoM of the star and planet. When orbits are labelled with $r$, we simply compute azimuthally-averaged components of the eccentricity vector along the azimuth. We then have a set of orbit-averaged elements as a function of radius $\{r,e(r),\varpi(r)\}$. We also obtain a simple azimuthally averaged surface density $\Sigma(r)$.
\item When labelling orbits using $a$ or $\lambda$, one has to be more careful. Let us define $d=a,~\text{or}~\lambda$. In order to characterize elliptical motion in the disc, we split the disc in bins of $\Delta d_0=[d_0-\delta d_0:d_0+\delta d_0]$. We then explore the grid and identify all cells for which $d$ lies in the interval $\Delta d_0$. That way, we have identified cells which share the same semi-major axis or semi-lactus rectum. We then average over these cells to compute orbit-averaged elements which are now labelled by $d$: $\{d,e(d),\varpi(d)\}$. We also obtained an averaged surface density $\Sigma(d)$, which is the mean of the density in each cells that share the same semi-major axis. 
\end{itemize}
On Figure \ref{fig:rsmaslr} we show the surface density, eccentricity and AMD as a function of $r$, $a$ or $\lambda$. Not surprisingly, there is very little difference between $a$ and $\lambda$ since the eccentricities are small. There are however noticeable differences between $r$ and $a$. In particular, the density profile shows a much larger gap in $a$. This will have an impact on the amplitude of the resonances, since they depend on the value of the surface density in the vicinity of the resonance. The eccentricity distribution is also shifted outward in $a$, resulting in a different distribution of AMD.

In the remaining of the paper, we use a labelling of orbits through their semi-major axis $a$.

\section{Results of numerical simulations}
\label{sec:results}

\subsection{Mesure of the growth rate}
As noted by \citet{kd06}, the radial kinetic energy $\Kr$ provides an easy way of measuring the eccentricity growth of the disc. It reads:
\begin{equation}
\label{eq:ekr}
\Kr = \int_{\rin}^{\rout} \int_{0}^{2\pi} \frac{1}{2}\Sigma u_{\rm r}^2 r \id r\id \theta,
\end{equation}
where $u_{\rm r}$ is the radial component of the velocity. We expect the eccentricity growth rate to be half that of the kinetic energy.

An other quantity of interest is the total angular momentum deficit (AMD) of the disc, which reads
\begin{equation}
\label{eq:amd}
A_{\rm d} = \int_{\rin}^{\rout} \int_{0}^{2\pi} \Sigma r^2\Omega (1-\sqrt{1-e^2}) r\id r\id \theta.
\end{equation}
In the linear regime, it scales as $e^2$ and should therefore grow on the same timescale as the radial kinetic energy. 

$\Kr$ presents the advantage that it is readily accessible from the variables output by the code. However we have seen that $u_{\rm r}$ (as measured in a frame centred around the star) will not correctly represent eccentric motion that would take place around the CoM. With most of the eccentricity growth taking place outside the orbit of the planet, $u_{\rm r}$ needs to be measured around the CoM. We then chose to measure the growth of eccentricity using the AMD of the disc. In equation (\ref{eq:amd}) we use quantities computed as described in Section \ref{sec:orb}. That is, the radial coordinate $r$ has to be interpreted as a semi-major axis.

\begin{figure}
    \begin{center}
    \includegraphics[scale=1]{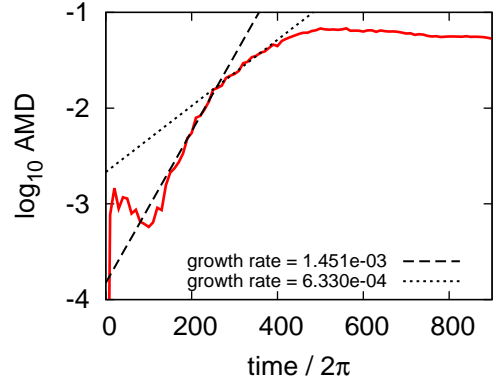}
    \caption{Growth rate (in units of $\Op$) measured from the disc total AMD for $\qp=7\times 10^{-3}$. We show two linear fits, illustrating the uncertainty in the measurement of the growth rate. The growth rate of eccentricity is obtained by multiplying the gradient by $\log(10)/(2\pi\times2)$ to correct for the $y$-axis scaling, time unit, and the fact that the AMD goes as $e^2$, respectively.}
    \label{fig:fiduamd}
    \end{center}
\end{figure}

In addition, as the planet opens a gap, strong tidal forces are exerted on the disc. As a consequence, cells in the vicinity of the planet might appear to have some spurious eccentricity. On Figure \ref{fig:fiduamd} we show the time evolution of the total AMD of the system. A violent increase is first observed over the first 10 orbits as the planet is injected into the disc. As the gap opens, the system slowly relaxed, until the exponential growth phase takes place and eventually saturates. Such behaviour was also observed by \citet{kd06} in the evolution of the radial kinetic energy. In principle it is possible that the eccentric mode starts growing earlier than what is shown on Figure \ref{fig:fiduamd}, on top of the tidal perturbation. In order to filter out the tidal perturbation, we only measure the growth of AMD between $a=2$ and $a=4$ in the remaining of the paper. 

\subsection{From simulations to secular theory}

Once we have measured the eccentricity distribution, growth rate and possibly precession rate from the simulations, we wish to compare it with the predictions of the linear theory. In order to compute eccentric modes from the secular theory, one needs to extract the surface density profile from simulations. As the disc evolves in time, the gap will become more eccentric, wider and more depleted, which will affect the results of the linear theory. On panel $b$ of Figure \ref{fig:intersigma}, one can see that the surface density indeed evolves in time. However, during the linear growth phase (between $t=200\tp$ and $t=400\tp$), the change in density remains small. The surface density profiles that we use in this paper are computed as follows: For each simulation, we identify the linear growth phase. We then compute a time-average density over a series of surface densities measured during the growth phase. The sampling is one point every 10 orbits, and we average over about 5 points. To satisfy the boundary conditions used in the linear theory, we have also reduced the surface density to zero over a few grid points at both edges of the disc. 

\begin{figure}
    \begin{center}
    \includegraphics[scale=1]{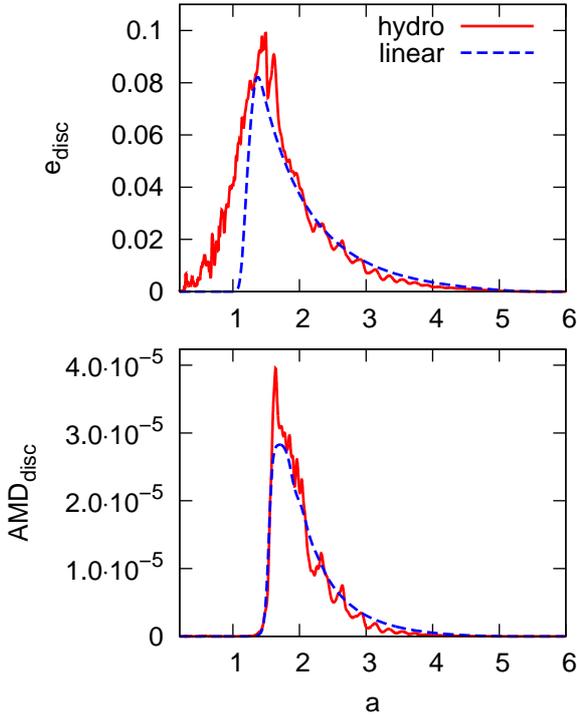}
    \caption{Distribution of eccentricity (top) and AMD (bottom) for $\qp=7\times10^{-3}$, from  direct hydrodynamical simulations (red solid line) and linear calculations (blue dashed line) as a function of semi-major axis $a$. The linear mode is scaled arbitrarily to match the eccentricity distribution for $a>2$. The amplitude of the AMD distribution is arbitrary since our simulations were carried with a surface density scaled by $\Sigma_0=1$.}
    \label{fig:qp7eamd}
    \end{center}
\end{figure}

This surface density profile can then be injected in the linear theory, and we then solve for the eccentric modes, as described in Section \ref{sec:secular}. On Figure \ref{fig:qp7eamd} we show the eccentricity distribution obtained from the hydrodynamical simulation, compared with the relevant eccentric mode obtained from the linear theory, for our fiducial example with $\qp=7\times 10^{-3}$. The scaling being arbitrary in the linear theory, we have scaled the mode so that its amplitude matches that of the simulation in the outer parts of the disc. Using the surface density, we can also plot the distribution of AMD in both cases. Overall, the simulations and linear theory show an excellent agreement regarding the shape of the mode in the outer parts of disc. Some eccentricity in the inner disc is observed in simulations, which is not described by this mode (although it is possible that another mode could grow in the inner disc, almost independently of the outer mode, due to the weak communication of eccentricity across the gap, but we do not attempt to study this mode in the present work). The simulation shows an excess of eccentricity between $a=1.4$ and $a=1.7$, which is not predicted by the linear theory. Interestingly, there is a corresponding strong departure from Keplerian orbits in this region, as can be observed in the lower panel of Figure \ref{fig:kappaom}. This could be an indication of the limitations of the linear theory in this regime. The two peaks are also located near the 2:4 and 3:5 ELRs. Once weighted by the surface density, this leaves one prominent peak in the AMD distribution. Apart from that peak, the distribution of AMD from the linear theory is in very good agreement with what was obtained in numerical simulations.

\subsection{Influence of the mass of the planet}

\begin{figure}
    \begin{center}
    \includegraphics[scale=0.7]{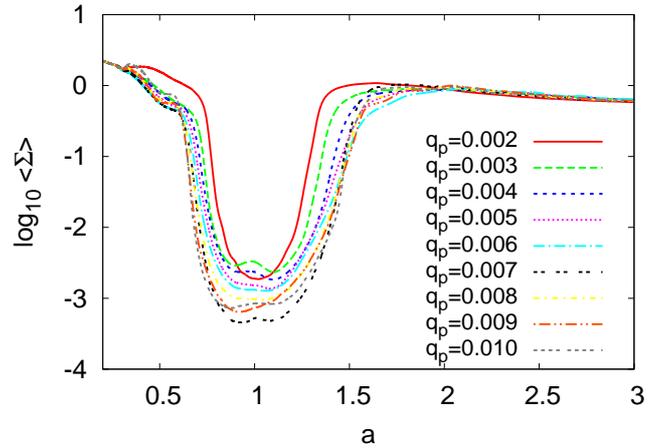}
    \caption{Orbit-averaged surface density profile as a function of the semi-major axis $a$ (zoomed between $a=0.2$ and $a=3$) for different planet-star mass ratios. These profiles are taken during the exponential growth phase of the eccentricity.}
    \label{fig:qp_sigma}
    \end{center}
\end{figure}

\begin{figure}
    \begin{center}
    \includegraphics[scale=0.7]{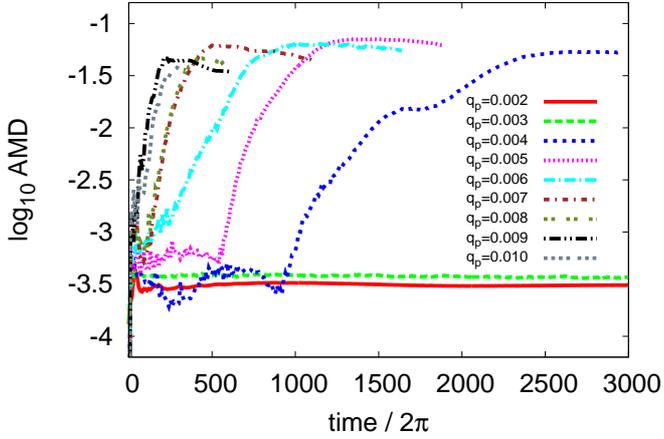}
    \caption{Growth rate of AMD as a function of time for different mass ratios $\qp$.}
    \label{fig:qp_amd}
    \end{center}
\end{figure}

\begin{figure}
    \begin{center}
    \includegraphics[scale=0.7]{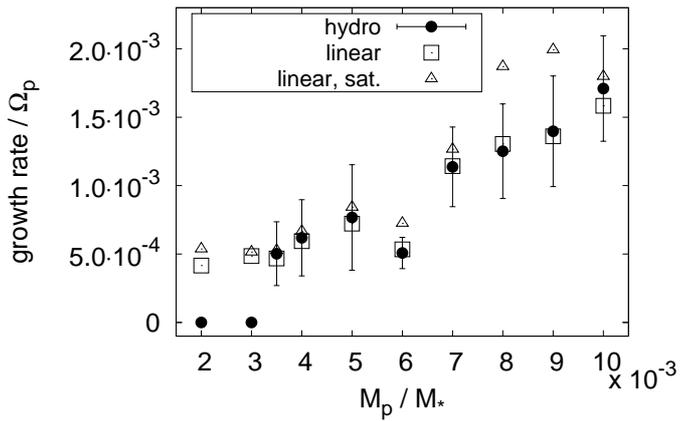}
    \caption{Eccentricity growth rate as a function of planet-to-star mass ratio. Full circles are results from direct numerical simulations. Empty squares indicate growth rates computed from linear calculations. Triangles are growth rates from the same linear calculations, assuming that the eccentric corotation resonances are fully saturated.}
    \label{fig:qp_growth}
    \end{center}
\end{figure}

For a given density profile, one would expect that more massive planets generate a larger growth rate in the disc, as the contribution from ELRs to the growth rate scales as $\Mp^2$. However, as the mass of the planet increases, and for a given viscosity, the gap will become progressively larger. This causes resonances to become progressively weaker, as the surface density in the vicinity of their resonant radius become more and more depleted (see Fig. \ref{fig:qp_sigma}). In addition, more massive planets lead to deeper gaps. The removal of material in the vicinity of the planet's orbit weakens the damping by eccentric co-orbital resonances. Finally, note that higher mass planets might also cause stronger departures from Keplerian orbits. Therefore one should not necessarily expect a smooth dependence of the growth rate on planet mass. In Figure \ref{fig:qp_amd} we show the evolution of AMD as a function of time for various $\qp$, and \ref{fig:qp_growth} we show the growth rate of eccentricity as a function of planet mass (black dots). 

We find that growth of eccentricity in the disc occurs only for $\qp> 3\times10^{-3}$. This is in agreement with the work of \citet{kd06}, although they find the transition to be between 2 and $3\times10^{-3}$. In the range $\qp=3.5-10\times10^{-3}$, the increase of growth rate with mass is almost linear (with the exception of a spurious point at $\qp=6\times10^{-3}$). In Figure \ref{fig:qp_growth} we also show the growth rates as computed from secular theory. The squares indicate the growth rates as computed when ECRs are fully operative, while crosses indicate the growth when ECRs are fully saturated. The difference between the two is small, showing that the saturation of ECRs does not play a major role in determining the growth of eccentricity. This result arises primarily from our modelling of ECRs which, in our treatment, operate on a narrower width than ELRs. Contrary to what our numerical experiments suggest, the linear theory predicts that gap-opening planets with $\qp \lesssim 3\times10^{-3}$ should cause the disc to become eccentric. However once the growth is observed in the simulations, its agreement with the linear theory is excellent.

\subsection{Influence of the viscosity}
\begin{figure}
    \begin{center}
    \includegraphics[scale=0.7]{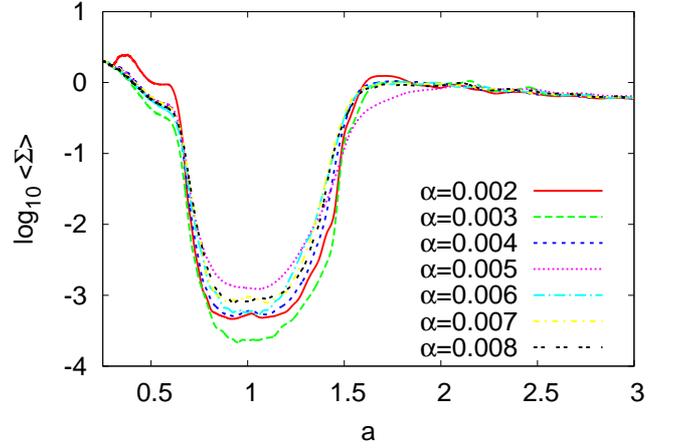}
    \caption{Orbit averaged surface density profile for different viscous $\alpha$ parameters. These profiles are taken during the exponential growth phase of the eccentricity.}
    \label{fig:visc_sigma}
    \end{center}
\end{figure}

In \citet{go06}, the influence that an effective bulk viscosity described by an $\alpha$-prescription would have on the eccentricity was derived. In this simple picture (motivated by our poor knowledge of angular momentum transport in accretion discs), a larger viscosity would cause a stronger damping of the eccentricity. However, in the case of gap-opening planets, viscosity determines the size of the gap by balancing the gravitational torque exerted by the planet. In this case, a larger viscosity means a narrower gap (for a given planet mass). A narrower gap means that more resonances can operate in a region of the disc that is not strongly depleted of material. If the net effect of all resonances is a growth of eccentricity, this indicates that higher viscosity can potentially mean higher growth rate. This effect is more subtle than the viscous damping, but the effect of viscosity on the gap width and depth can clearly be seen in Figure \ref{fig:visc_sigma}. The viscous parameter also causes the width of the ECRs to become wider at larger viscosity (see equation \ref{eq:rwecrv}), which should increase the damping.

\begin{figure}
    \begin{center}
    \includegraphics[scale=0.7]{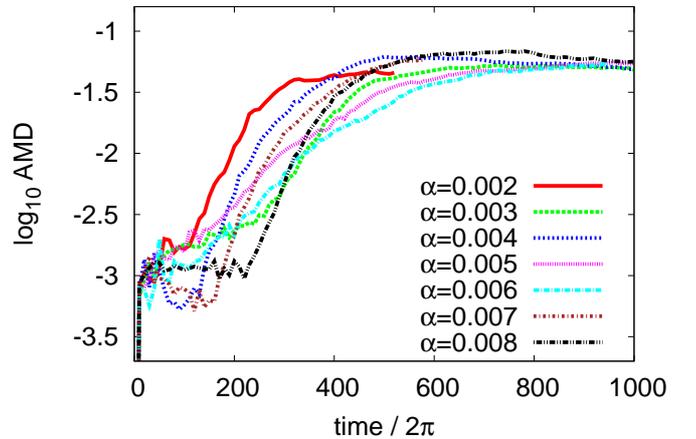}
    \caption{Growth rate of AMD as a function of time for various values of the viscous parameter $\alpha$.}
    \label{fig:visc_growth}
    \end{center}
\end{figure}

The growth of AMD in the disc is shown in Figure \ref{fig:visc_growth}. It shows a non-monotonic behaviour of the growth rate with viscosity, but all runs seem to eventually saturate at the same value of AMD. We have computed the growth rate predicted from the linear theory, shown in Figure \ref{fig:visc_growth}. In several cases the linear theory fails to reproduce the growth rate observed in simulations. However, the linear theory captures the same non-monotonic behaviour of growth rate with viscous parameter. Capturing the effect of viscosity in both numerical simulations and the linear theory is a tricky problem, and it worth reminding that the numerical simulations use a shear viscosity, while the linear theory uses a bulk viscosity. In any case, the actual contribution from the viscous term to the damping of eccentricity is small compared to the growth and damping caused by the ELRs and ECRs, respectively, which we discuss in the next section.

\begin{figure}
    \begin{center}
    \includegraphics[scale=0.7]{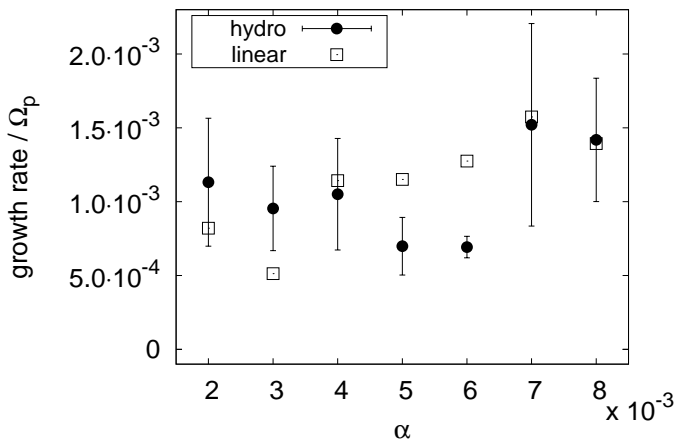}
    \caption{Eccentricity growth rate as a function of the viscous $\alpha$ parameter. Full circles are results from direct numerical simulations. Empty squares indicate growth rates computed from linear calculations.}
    \label{fig:ekrad_pf}
    \end{center}
\end{figure}

\subsection{Contribution from single corotation and Lindblad resonances}

\begin{figure}
    \begin{center}
    \includegraphics[scale=0.65]{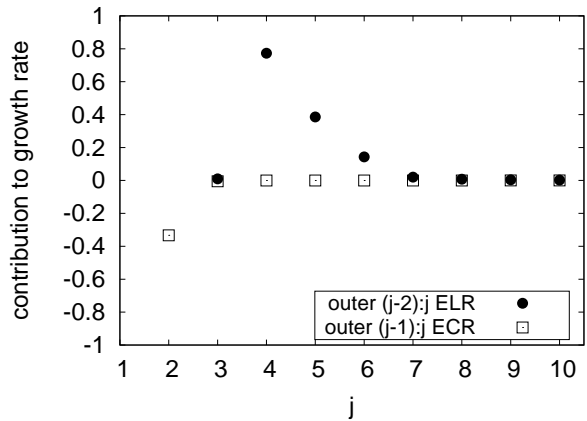}
    \caption{Contribution from various resonances to the growth rate of the fastest growing mode for $\qp=7\times 10^{-3}$. Resonances in the outer disc only are shown for simplicity. Contributions are normalized relative to the total growth rate of the mode. For simplicity, we identify ELRs with $(j-2)/j$ mean-motion resonances, and ECRs with $(j-1)/j$ mean-motion resonances. The 2:4, 3:5 and 4:6 ELRs contribute to most of the growth rate, with a significant damping from the 1:2 ECR. Resonances are computed up to $j=r/H=20$, but the contribution for all $j>10$ is negligible and not represented here.} 
    \label{fig:qp_resstrength}
    \end{center}
\end{figure}

On Figure \ref{fig:qp_resstrength} we show the contribution to the growth rate from various Lindblad and corotation resonances, as obtained from the linear calculations. As in \citet{dlb06} and \citet{to16}, we find that most of the contribution to the growth rate comes from the 2:4, 3:5 and 4:6 ELRs, with the 1:2 ECR giving the only significant contribution to damping. In particular the 1:3 ELR is not key in determining the growth rate of eccentricity in the disc. Even without the saturation of the 1:2 ECR, the contribution of all ELRs still gives a net growth of eccentricity.

\subsection{Precession rate}

The real part of the eigenfrequency of the mode of interest here represents the precession rate the eccentric mode. For the case where $\qp=7\times 10^{-3}$, the linear theory predicts a prograde precession with $\Re(\omega)=7.71\times10^{-4}\Op$. For 2D discs, we showed in \citet{to16} that the pressure contributes to the precession via two terms, one leading to retrograde precession, and the other to prograde or retrograde, depending on the pressure gradient. In addition, the secular disc-planet interaction leads to prograde precession of the mode (see equations \ref{eq:ip1}, \ref{eq:ip2} and \ref{eq:ipd}). In our case, we find the contribution of these three terms to be $I_{\rm p1}= -2.99\times 10^{-3}\Op$, $I_{\rm p2}= 2.67\times 10^{-3}\Op$ and $I_{\rm pd}=1.08\times 10^{-3}\Op$ respectively. Therefore the two terms due to pressure, $I_{\rm p1}$ and $I_{\rm p2}$ nearly cancel each others, and most of the contribution comes from the prograde precession driven by the secular forcing from the planet, $I_{\rm pd}$.

\begin{figure*}
\centering
\begin{minipage}{.48\textwidth}
\centering
\includegraphics[width=\linewidth]{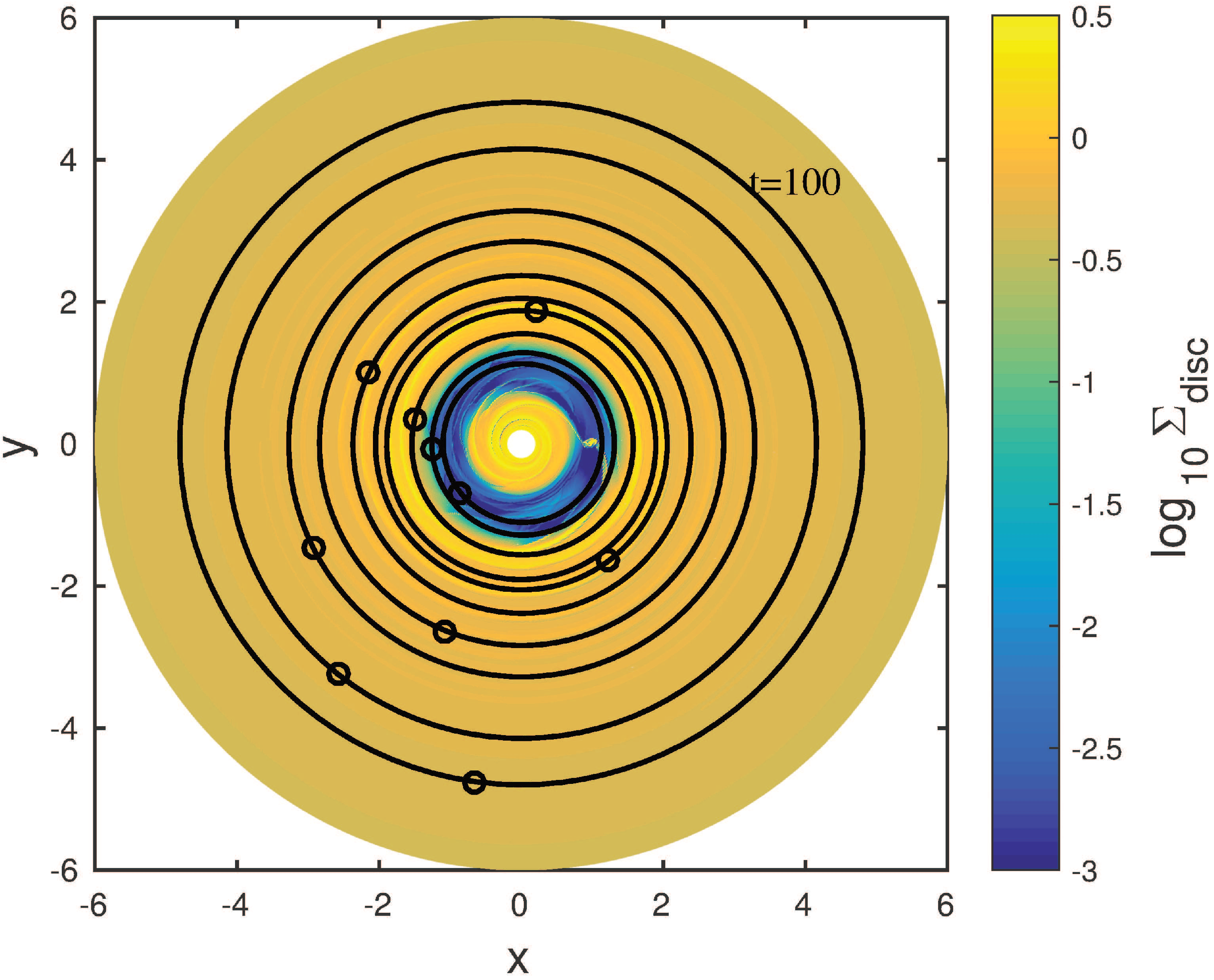}
\label{fig:test1}
\end{minipage}\hfill
\centering
\begin{minipage}{.48\textwidth}
\centering
\includegraphics[width=\linewidth]{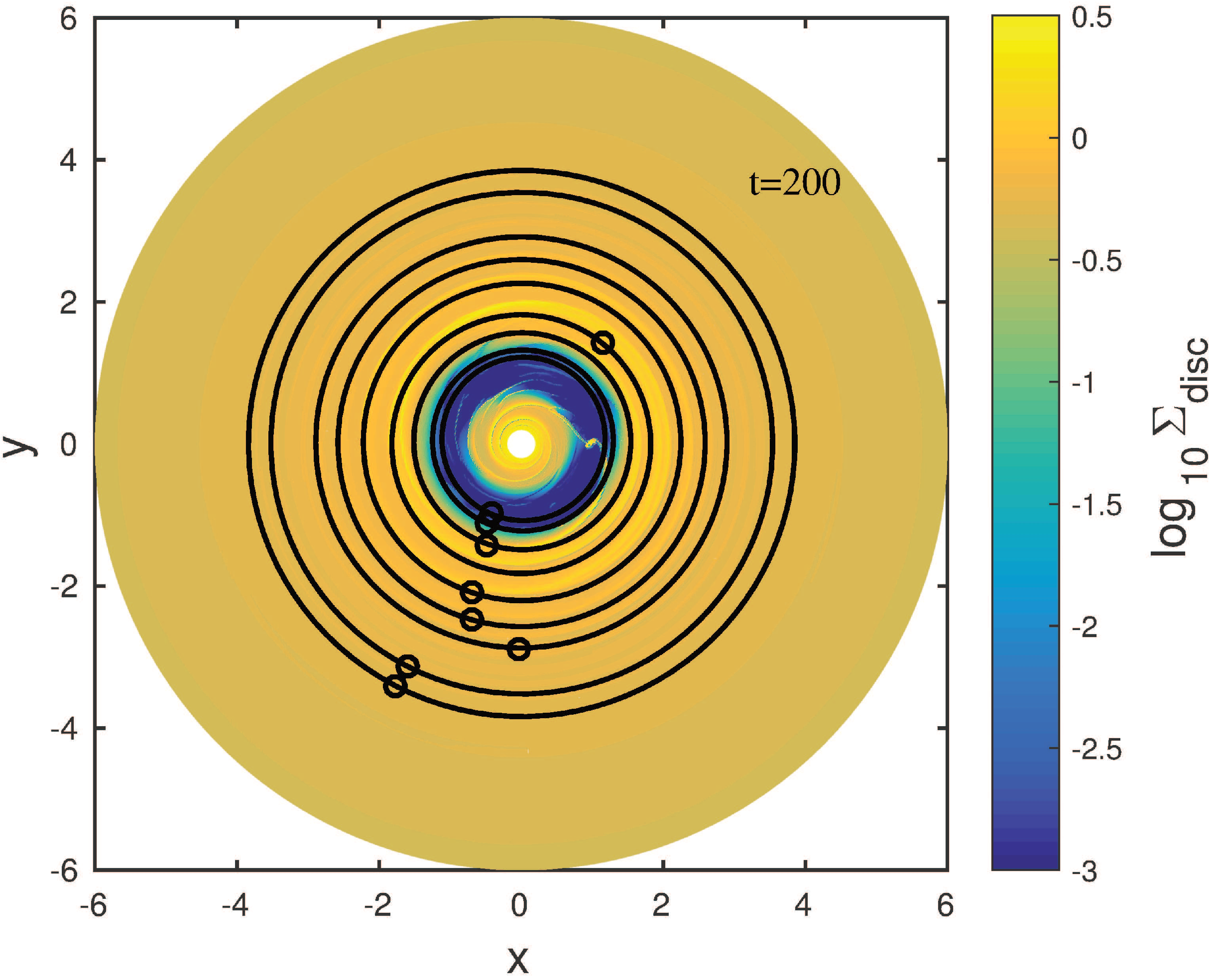}
\label{fig:test1}
\end{minipage}\hfill
\begin{minipage}{.48\textwidth}
\centering
\includegraphics[width=\linewidth]{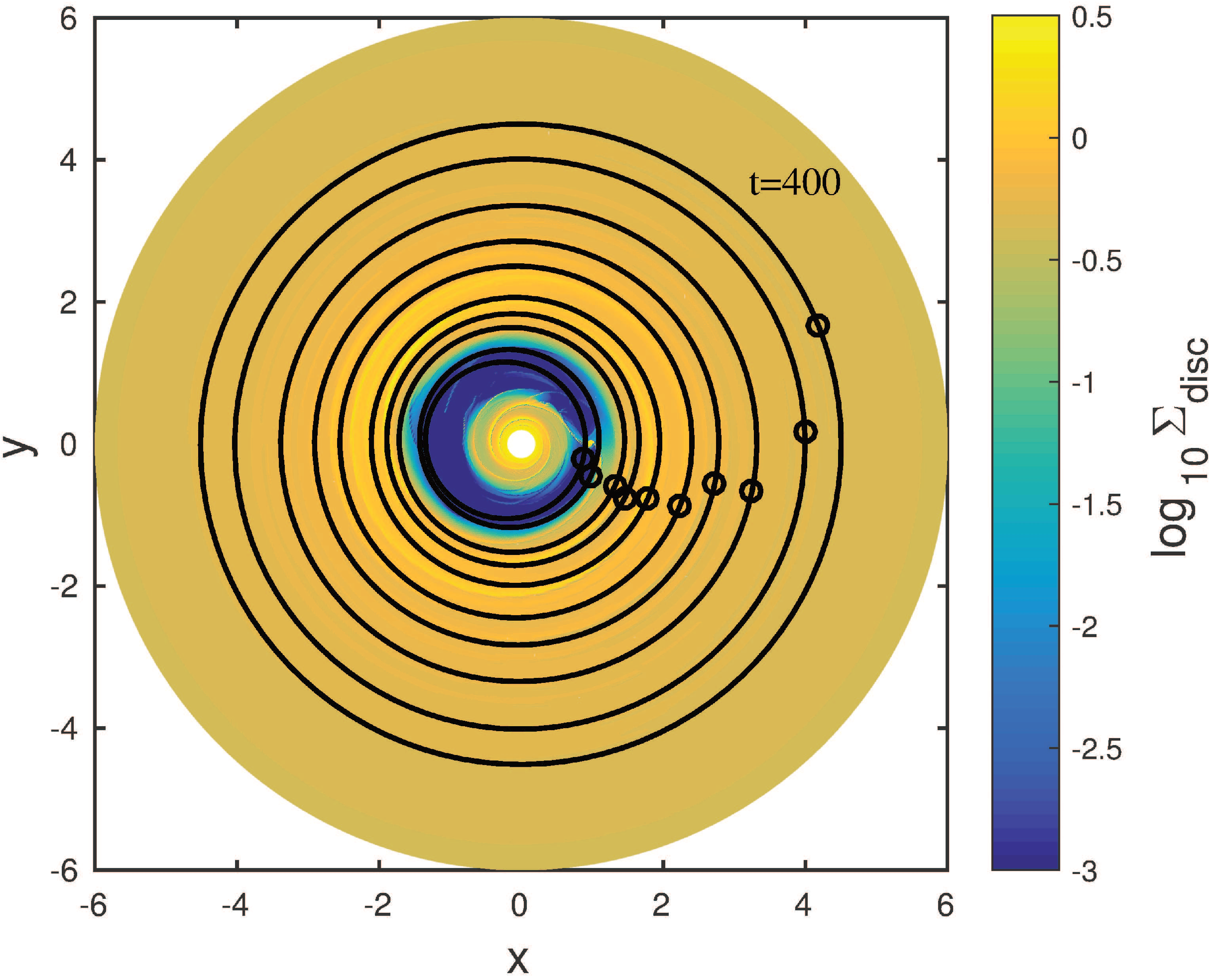}
\label{fig:test2}
\end{minipage}\hfill
\begin{minipage}{.48\textwidth}
\centering
\includegraphics[width=\linewidth]{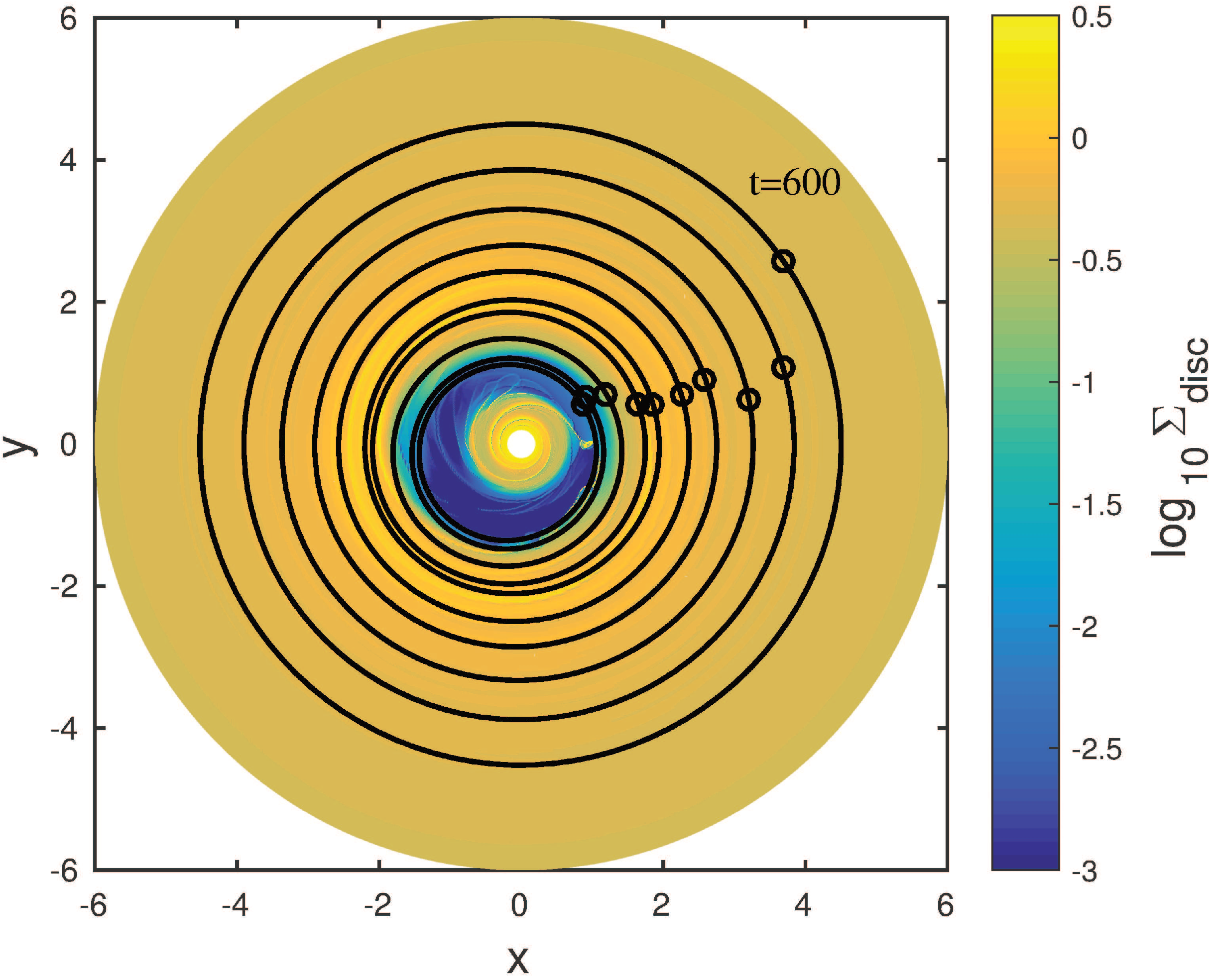}
\label{fig:test3}
\end{minipage}
\caption{Surface density of a disc with  $\qp=7\times 10^{-3}$, at times $t=100$, 200, 400, 600$\tp$ (see Fig. \ref{fig:fiduamd} for the corresponding stage in the evolution of the system). Also shown are various ellipses representing eccentric rings fitted to the velocity field of the disc. The black circle represents the pericentre of the ring. Prograde precession is observed, and the mode precesses more and more coherently with time.}
\label{fig:precdisc}
\end{figure*}

\begin{figure}
    \begin{center}
    \includegraphics[scale=0.7]{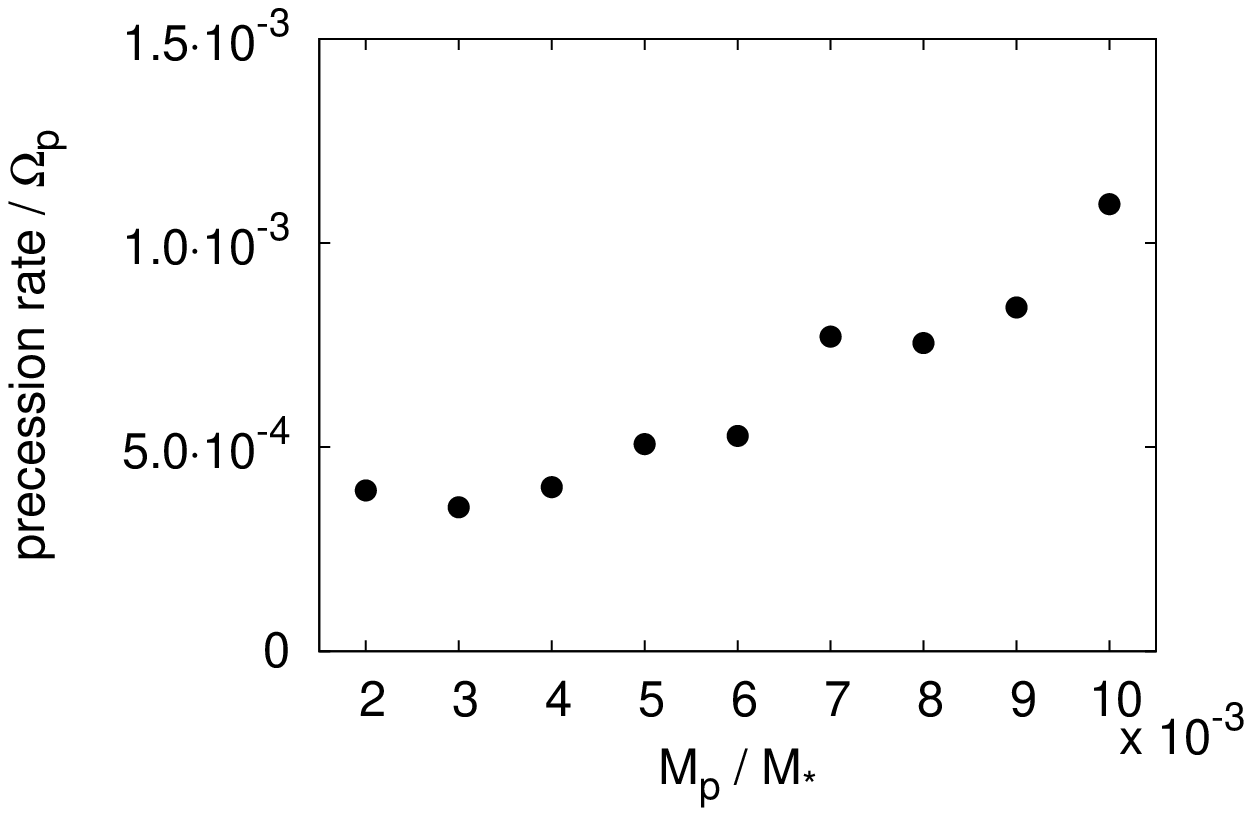}
    \caption{Eccentricity precession rate as a function of planet-to-star mass ratio, from the linear calculations only.}
    \label{fig:qp_prec}
    \end{center}
\end{figure}

On Figure \ref{fig:precdisc} we show a disc at different snapshots for the same mass ratio. Plotted on top of the surface density are a set of Keplerian orbits, where the dot indicates the pericentre. These snapshots clearly indicate a prograde precession of the mode. At $t=100\tp$, the system is not yet in the linear regime yet (see Fig. \ref{fig:fiduamd}), and there is no coherent precession. A coherent precession starts to appear in the inner part of the outer disc, in the linear phase at $t=200\tp$ and after, although with a slight twist in the outer part of the disc. This twist could arise because the mode is growing faster tan it is precessing, and has not have the time to reach the outer part of the disc while it develops in the inner part. Note also that in the outer disc, the eccentricities are so small that it is hard to define a numerically accurate pericentre. This twist makes it hard to measure an accurate precession rate for the mode. However one can see that from $t=200\tp$ to $t=400\tp$, the mode has precessed by about 60 degrees. This corresponds to a precession rate of $\Re(\varpi)=8.33\times10^{-4}\Op$, which is in broad agreement with the linear theory. On Figure \ref{fig:qp_prec} we plot the precession rate as obtained from the linear theory. Due to the increasing effect of the gravitational interaction with increasing planet mass, the precession rate increases with mass.

\subsection{Saturation of the eccentricity}

\begin{figure}
    \begin{center}
    \includegraphics[scale=0.7]{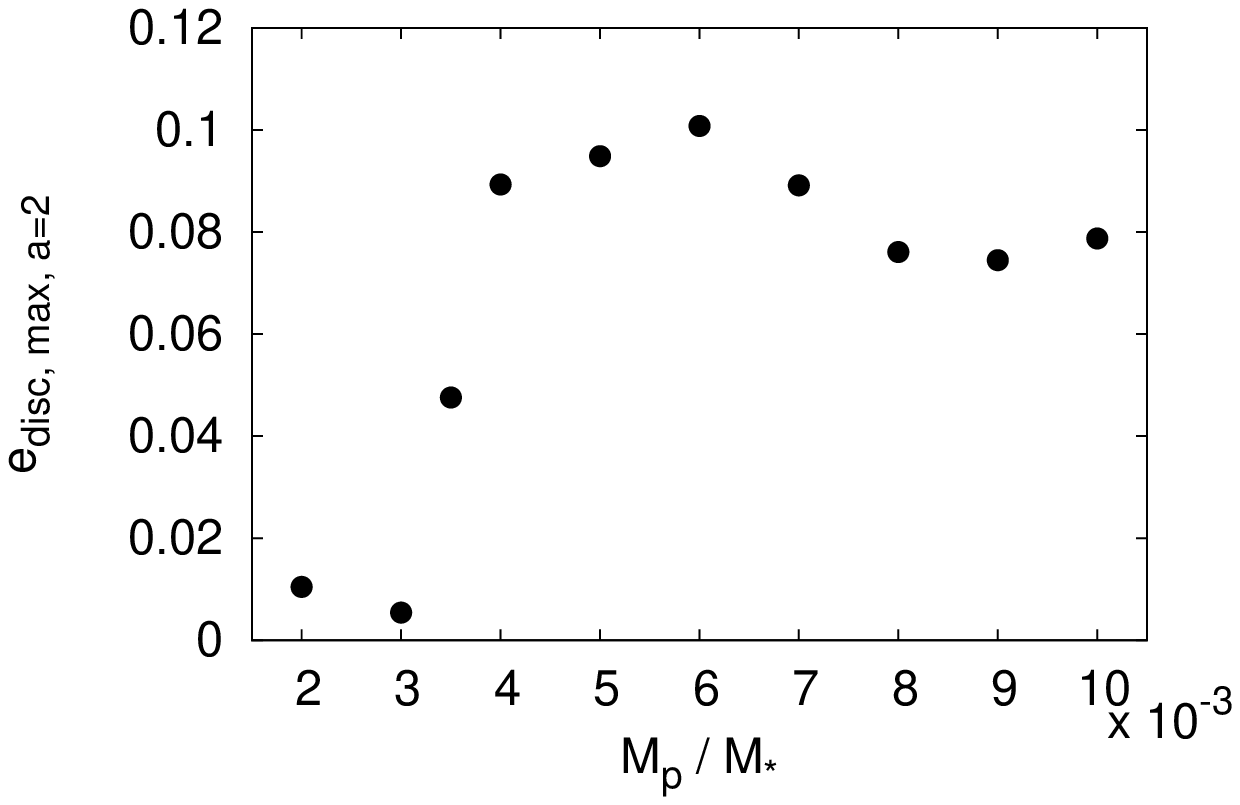}
    \caption{Maximum ortbit averaged eccentricity (squares, left $y$-axis) and AMD (circles, right $y$-axis) reached by the disc at a semi-major axis $a=2$, as a function of planet-star mass ratios.}
    \label{fig:qp_emax}
    \end{center}
\end{figure}

\begin{figure}
    \begin{center}
    \includegraphics[scale=0.7]{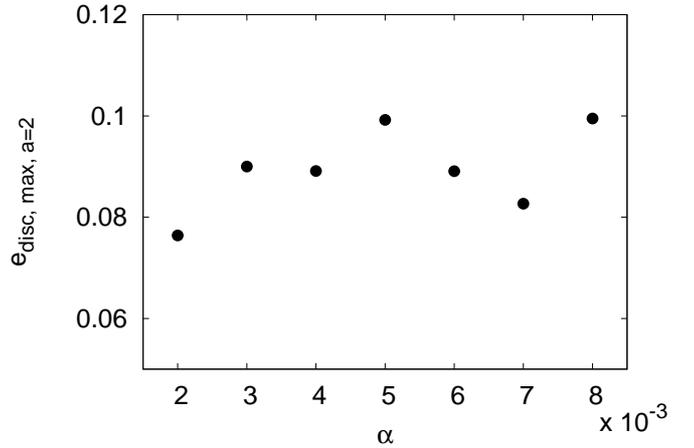}
    \caption{Maximum orbit averaged eccentricity (squares, left $y$-axis) and AMD (circles, right $y$-axis) reached by the disc at a semi-major axis $a=2$, as a function of the viscous $\alpha$ parameter.}
    \label{fig:visc_emax}
    \end{center}
\end{figure}

The saturation of eccentricity seen in the simulations is not accounted for by the linear theory. On Figures \ref{fig:qp_emax} and \ref{fig:visc_emax} we show the maximum value reached by the eccentricity of the disc as a function of the planet's mass and viscosity, respectively. We evaluate the maximum eccentricity at a semi-major axis of $a=2$. The eccentricity might be larger than this value elsewhere in the disc, such as in the gap. However at $a=2$ the AMD is larger than in the gap, and agrees very well with the linear theory, so we chose it at a robust marker of the maximum eccentricity. The transition from no growth to growth between $\qp=3\times10^{-3}$ and $\qp=4\times10^{-3}$ is clearly visible in Figure \ref{fig:qp_emax}. Interestingly, we have conducted an additional run with $\qp=3.5\times10^{-3}$, which showed eccentricity growth after 3000 orbits, but saturated at an intermediate value of $e=0.048$ (see Figure \ref{fig:qp_emax}), suggesting a smooth transition between the two regimes. Its growth rate agrees well with that obtained from the secular theory. Finally, the maximum eccentricity does not show a clear dependence on viscosity (Figure \ref{fig:visc_emax}). 

\begin{figure*}
    \begin{center}
    \includegraphics[scale=1.25]{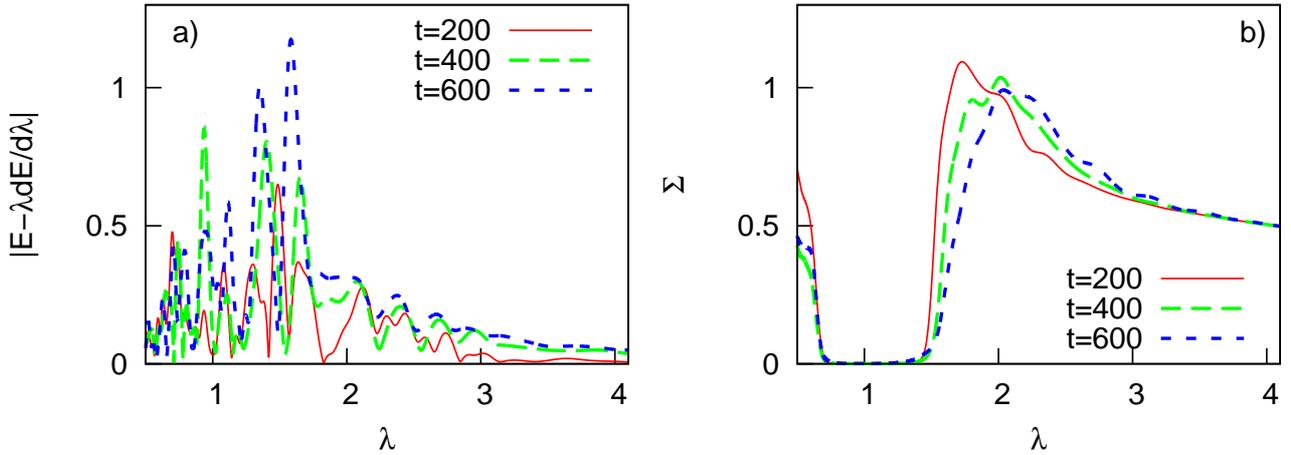}
    \caption{\textit{left:} Orbit intersection criterion as a function of the semi-latus rectum $\lambda$ (zommed between 0.5 and 4) for $\qp=7\times 10^{-3}$, at $t=200, ~400 \text{ and }600 \tp$. In the non-linear regime, orbital intersection is taking place in the disc, which can limit the eccentricity growth. \textit{right:} Surface density of the disc at the same times. In the non-linear regime, the gap has become wider, reducing the strength of resonances.}
    \label{fig:intersigma}
    \end{center}
\end{figure*}

We now turn to why the eccentricity saturates. The linear theory predicts an exponential growth, and non-linear effects must be taking place to halt this growth. The first non-linear effect could be the consequence of near-intersecting orbits. In the linear theory, orbits must be nested and not intersect. \citet{ogilvie01} showed that orbit intersection will take place when $|E-\lambda \id E/ \id \lambda|\approx 1$. On the left panel of Figure  \ref{fig:intersigma} we show this quantity as a function of $\lambda$ for our fiducial example, at 3 different times. At $t=400\tp$, towards the end of the eccentricity growth phase, $|E-\lambda \id E/ \id \lambda|$ starts becoming significant in parts of the disc. At $t=600\tp$, it has reached values around 1 in at least two locations in the disc. Interestingly, these two locations could correspond to the peaks observed in the eccentricity distribution in Figure \ref{fig:fiduamd}. Their location could also correspond to the 2:4 and 3:5 ELRs, where eccentricity excitation is likely to be the most important.

In the right panel of Figure \ref{fig:intersigma} we show the orbit-averaged surface density for the same simulation, at the same times. At later times, the gap becomes larger, which reduces the strength of the resonances. This can also reduce the eccentricity growth. 

We have not investigated other effects such as non-linear resonances, that could also play a role when the eccentricity in the disc becomes significant. Further work on the subject is needed but in the light of the present paper we favour orbital intersection  as a mechanism to limit eccentricity growth. 

\section{Discussion}
\label{sec:discussion}

In this paper we have developed a framework for the study of eccentric discs. We have highlighted the benefit of describing the disc using a set of Keplerian orbits instead of a simple azimuth-averaged method. This has consequences for the surface density profile and shape of the eccentric mode. We have also shown how the secular theory of eccentricity for planet-disc interactions can reproduce some of the main features observed in simulations, such as the shape of the mode in the outer disc, the distribution of AMD, and the growth rate and precession rate of the disc. Numerically, we have recovered the result of \citet{kd06} that planet-to-star mass ratios larger than $\qp=3\times10^{-3}$ can lead to eccentricity growth in the disc. However several assumptions were made, and some discrepancies remain between the linear theory and the simulations, which we discuss in this Section.

In this paper we have kept the planet on a fixed circular orbit for simplicity. The results of  \citet{to16} suggested that there is a good separation of the timescales on which the eccentricity of the disc and the planet grow under resonant effects, with the disc-dominated timescale being shorter than the planet-dominated timescale. Evidence in support of this separation of timescales was observed by \citet{rbctfm16} who found the growth rate of a $10$ Jupiter-mass planet to be of the order of $10^{-7}\Op$ (a result which will of course depend on the disc mass). There is however an intermediate timescale which might have led to misinterpretation in the literature, and which is set by the secular exchange for angular momentum between the disc and the planet. These secular variations will lead to a periodic variation of eccentricity. Suppose that the timescale associated with these variation is $10^{4}~\tp$. A simulation carried over a time which is half this period might indicate a growth (or decay) of the eccentricity of either the disc or the planet, while it is in fact a secular oscillation that is not related to any growth mechanism. The results of \citet{rbctfm16} clearly show a net growth of eccentricity on top of secular variations.

In \cite{to16} we have discussed the major role played by a term arising from 3-dimensional effects, which would help trapping a mode in the inner parts of the disc, where it could rapidly grow. This 3D term was not included here since the aim of this paper was a comparison with 2D simulations. In the present case, it is possible that the confinement of the mode is made possible by the potential well created by the gravitational field of the planet and by the choice of outer boundary condition that forces the mode to go to zero at the outer edge. The AMD of the mode is therefore reduced, causing the mode to grow more rapidly than if it was not confined. Without this boundary term, the possibility for the mode to be trapped would depend on the choice of outer radius. An other mechanism by which the mode could be trapped is due to the fact that it is growing faster than it is precessing. In other words, the resonant interactions cause the mode to grow before it can reach the outer parts of the disc. In this case the trapping of the mode would be a local effect due to resonances. This is supported by the twist in the arguments of pericentre seen in Figure \ref{fig:precdisc}, which could be the result of a mode that has grown rapidly in the inner part of the disc while it has not fully developed in the outer disc yet. In any case, the inclusion of the 3D term would certainly help the trapping of a mode in the inner disc. If the 3D effect were to be important, it would only enhance the growth rate of eccentricity in the disc. In addition, the presence of a deep gap significantly reduces the communication of eccentricity between the inner and outer discs via pressure effects. Note that in massive discs, self-gravity (which we did not include here) would be another way for the disc to communicate eccentricity across the gap. 

The present work assumed the disc to be locally isothermal. The advantages of these assumptions are twofold: from the numerical point of view, it provides a fairly simple set-up, and can be compared with previous work of the same kind. From the secular theory point of view, the formalism was already developed in \citet{to16}, and was straightforward to apply to the present work. One caveat is that in the secular theory, locally isothermal discs present the inconvenience of not conserving AMD (although a related quantity has conservation properties in simple cases, see Appendix \ref{app:int}). When going beyond the assumption of locally isothermal discs, it is clear that thermal effects play an important role in protoplanetary discs. A secular theory allowing for thermal effects in the disc would certainly shed new light on the eccentricity evolution of protoplanetary discs. In addition, \citet{ttc14} have discussed the possibility for the eccentricity of a planet in a gap to grow when the gap is illuminated by stellar irradiation, which modifies the entropy gradient across the gap, and therefore affects the corotation torque. 

The question remains why no eccentricity growth is observed in simulations with mass ratios of $\qp\lesssim 3\times10^{-3}$. We have run the simulations of $\qp=2\times10^{-3}$ and $\qp=3\times10^{-3}$ for more than 7000 orbits and did not observe a growth. Extrapolating from the e-folding time for $\qp=3.5\times10^{-3}$ and $\qp=4\times10^{-3}$ and based on the prediction of the secular theory, it should have been sufficient to observe the growth. 
In Appendix \ref{app:num} we explore different numerical setups but find a good convergence. In Appendix \ref{app:resw} we show that in principle, our resolution was sufficient to resolve the narrow width of ECRs, and in particular the width of the strong 1:2 ECR is resolved. It is also possible that the linear theory fails to reproduce the correct growth rate at low $\qp$. One possibility would be the additional damping due material co-orbiting with the planet. The depletion of the gap is however expected to make such damping rather weak. In addition, Figure \ref{fig:qp_sigma} indicates that the gap for $\qp=2\times10^{-3}$ and $\qp=3\times10^{-3}$ is not less depleted than for $\qp=4\times10^{-3}$, therefore damping should be observed there too. The precise effect of each resonances depends strongly on the model we use. In particular, it might be possible that our treatment of ECRs leads to an underestimation of its damping. \citet{ol03} showed that saturation of the corotation torque is easier at large $\qp$. Therefore, one could envision a scenario in which our large $\qp$ simulations show saturation, while the low $\qp$ ones do not. Any error on the corotation resonance would then have a much bigger consequence at low $\qp$. One possible avenue of future work would be to consider the precise effect of the entropy gradient on the corotation resonances. As the rotation profile is not quite Keplerian, the vortensity gradient would need to be computed accordingly. Further complications arise for locally isothermal discs, where an additional component to the corotation torque is expected, that scales with the radial gradient of temperature. Such component is absent from the adiabatic case \citep[studied, e.g., by][]{bm08,tsang14}, and deserves a careful study. We have conducted two globally isothermal simulations (which do not suffer from this additional term) for $\qp=3\times10^{-3}$ and $\qp=7\times10^{-3}$. The simulation with $\qp=3\times10^{-3}$ did not show a growth of eccentricity while the one with $\qp=7\times10^{-3}$ did.
Finally we note that we have observed a growth of eccentricity for $\qp=3\times10^{-3}$ with a disc aspect-ratio of 0.025.

We conclude by pointing that more work is needed to understand the subtle mechanisms that lead to growth or decay of eccentricity in planet-disc interactions. Very long term simulations like those presented in \citet{rbctfm16} are necessary to study the long-term growth of eccentricity in systems where both the planet and disc are allowed to develop eccentricity. Because of the very long timescale on which the eccentricity of the planet grows, only such simulations carried over several secular periods can help disentangle between the net growth caused by resonant effects and the period variations caused by secular effects. Although it is unlikely that planet-disc interactions alone can explain the broad distribution of exoplanet eccentricities, it remains to be explored whether it can provide a seed of eccentricities (or more generally, a seed of AMD) that can serve as initial conditions for the onset of dynamical interactions between planets once the disc has been cleared away. In addition, three-dimensional effects can have a strong influence on the eccentricity growth \citep{ogilvie08,to16}, and should be studied in details, or at least be incorporated self-consistently in two-dimensional simulations. Finally, non-isothermal effects remain to be studied, both in the context of a linear theory and in simulations.

\section*{Acknowledgements}

We thank Quentin Andr\'e, Richard Booth and Giovanni Rosotti for useful discussions, and the referee for a constructive report. We acknowledge support from STFC through grant ST/L000636/1.

\bibliographystyle{mnras}
\bibliography{biblio2}

\appendix

\section{Supplementary information regarding the numerical methods}

\subsection{Convergence tests}
\label{app:num}

\begin{figure}
    \begin{center}
    \includegraphics[scale=0.6]{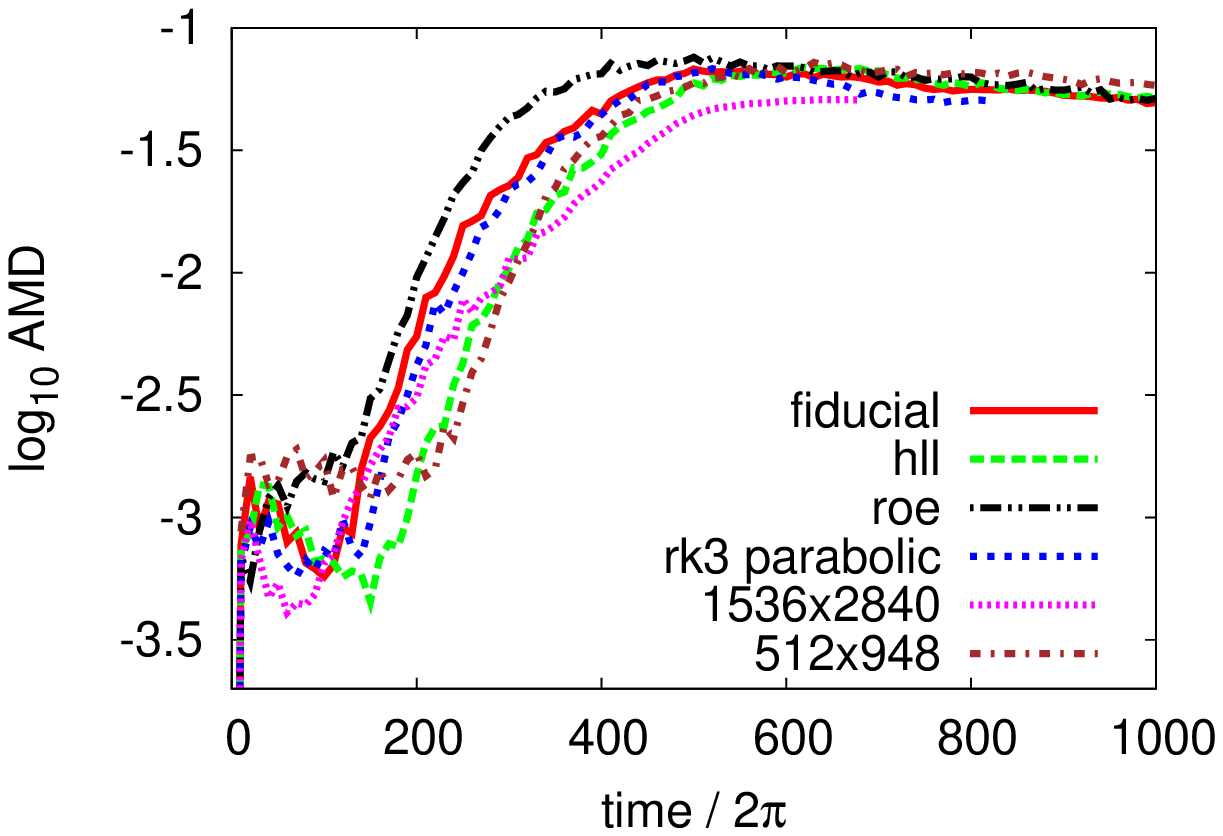}
    \caption{Convergence test for $\qp=7\times 10^{-3}$.}
    \label{fig:convtest}
    \end{center}
\end{figure}

In this appendix we give more information regarding the numerical tests we have done to check the consistency of our results. On Figure \ref{fig:convtest} we compare our fiducial $\qp=7\times 10^{-3}$ with five other simulations, all using \textsc{Pluto}. One run was conducted using the \textit{hll} solver, one run was conducted using the \textit{roe} solver, one using a third-order Runge-Kutta method with piecewise parabolic reconstruction, and two using different resolutions, $512\times 948$ and $1536\times 2840$. All runs consistently show a growth on a similar timescale and a saturation at the same value, although the high resolution run shows a slight decrease of the growth rate at later times. Note that the \textit{hllc} solver we used for this paper qualitatively reduces to an \textit{hll} solver in \textsc{Pluto} for a locally isothermal disc, since there is no contact discontinuity in this case.

\subsection{Resonance width}
\label{app:resw}
It is important to correctly resolve the resonance width, both for ELRs and ECRs. For the disc exterior to the planet, with our fiducial resolution in $r$, $N_{\rm r}=768$ on a log grid extending from $r=0.2$ to $r=6$, the resolution is $\Delta r/r\approx 4.4\times 10^{-3}$. The width of each resonance depends on its wavenumber $m$. For $m$ between 2 and 10, the width of ELRs  decreases from $w_{\rm L}/r_{\rm res}=6.5\times 10^{-2}$ to $4.5\times 10^{-2}$ with $H/r=0.05$.
For $\alpha=4\times 10^{-3}$, the width of ECRs decreases from $w_{\rm C}/r_{\rm res}=1.7\times 10^{-2}$ to $1.0\times 10^{-2}$ for $m$ between 1 and 10. Therefore large-$m$ ECRs are resolved over less than three grid points. However they are expected to be weak as they occur in the deep gap. The 1:2 ECR, which is the most important one, is correctly resolved.

\section{Secular theory of eccentricity in disc-planet systems}

\subsection{Discretized evolutionary equation}
\label{app:discretize}
We multiply equation (\ref{eq:2diso}) to (\ref{eq:ebc}) by $-\mi 2r/\cst$ to get the following equation for the evolution of eccentricity in a 2D isothermal disc:
\begin{align}
\label{eq:full2Diso}
-\mi\frac{2\Sigma r^3\Omega}{\cst}\dd{E}{t} & = \dd{}{r}\left[\Sigma r^3(1-\mi\alpha)\dd{E}{r} \right] \nonumber \\
&+\left[C(r)-\mi D(r)\right]E - \mi\alpha\frac{\Sigma r^3}{\cst}\dd{\cst}{r}\dd{E}{r},
\end{align}
where
\begin{equation}
C(r) = \text{pressure terms + secular gravitional terms}
\end{equation}
\begin{align}
D(r) & = \text{Resonant terms + viscous term} \\
     & +\text{boundary condition terms}.
\end{align}
The last term in eq. (\ref{eq:full2Diso}) is a non-adiabatic term. We now assume that the disc is made of a collection of $n$ non-intersecting annuli, with $i=1,2,\dots,n$, such as annulus $i$ occupies the interval $r_{i-1}<r<r_i$. We associate a unique eccentricity $E_i$ to each of these annuli. We multiply eq. (\ref{eq:full2Diso}) by $2\pi$ and integrate from $r_{i-1}$ to $r_i$ to get
\begin{equation}
-\mi J_i \dd{E_i}{t} = g_iJ_iE_i + 2\pi\left[F\dd{E}{r} \right]_{r_{i-1}}^{r_i} - \mi\dd{E_i}{r}V_i.
\end{equation}
Here we have introduced the following notations:
\begin{equation}
J_i = \int_{r_{i-1}}^{r_i} \frac{2\Sigma r^2 \Omega}{\cst}2\pi r \,\id r, 
\end{equation}
\begin{equation}
g_iJ_i = \int_{r_{i-1}}^{r_i} \left[C-\mi D \right]2\pi \,\id r,
\end{equation}
\begin{equation}
V_i =  \int_{r_{i-1}}^{r_i}\alpha  \frac{2\Sigma r^2}{\cst}\dd{\cst}{r}2\pi r \,\id r,
\end{equation}
and
\begin{equation}
F=\Sigma r^3 (1-\mi\alpha).
\end{equation}
Using forward finite differences for the radial derivatives, and assuming normal modes of the form $E=E(r)\me^{\mi\omega t}$, the discretized equation takes the form
\begin{align}
\omega J_i E_i & =  g_iJ_iE_i \nonumber \\
&+ 2\pi\left[\frac{F_{i}E_{i+1}}{\delta r_{i}} - \frac{F_{i}E_{i}}{\delta r_{i}} - \frac{F_{i-1}E_{i}}{\delta r_{i-1}} + \frac{F_{i-1}E_{i-1}}{\delta r_{i-1}} \right] \nonumber \\
&-\mi \frac{E_{i+1}V_i}{\delta r_i} +\mi \frac{E_iV_i}{\delta r_i}.
\end{align}

\subsection{Integral relations}
\label{app:int}

In this section we present useful expressions for the contribution to the precession rate and growth rate of various mechanisms. All integrals are carried from the inner to the outer radius of the disc. We start by defining the following quantity:
\begin{equation}
B = \int \frac{\Sigma r^2 \Omega}{2\cst}|E|^2  2\pi r \,\id r.
\end{equation}
The expression inside the integral is the AMD at the given radius in the disc, divided by the sound speed square. As shown in \citet{to16}, $B$ is conserved in inviscid, non-self-gravitating disc when resonances are not taken into account.

When looking for solutions for the eccentricity in the form of a normal mode $E(r)\me^{\mi \omega t}$, the precession rate of an eccentric mode is then $\Re(\omega)$, and the growth rate is $-\Im(\omega)$. We have the following relation for the precession rate:
\begin{equation}
\Re({\omega})=\frac{I_{\rm p1} + I_{\rm p2} + I_{\rm pd} + I_{\rm na}}{2B},
\end{equation}
where
\begin{equation}
\label{eq:ip1}
I_{\rm p1} = -\int\frac{1}{2}\Sigma r^2\left|\pd{E}{r}\right|^2 2\pi r \, \id r,
\end{equation}
\begin{equation}
\label{eq:ip2}
I_{\rm p2} = \int\frac{1}{2}\left[\frac{r^2}{\cst}\dd{\Sigma\cst}{r}- \frac{1}{\cst}\dd{}{r}\left(\Sigma \dd{\cst}{r}r^3 \right) \right]\left|E\right|^2 2\pi r \, \id r,
\end{equation}
\begin{equation}
\label{eq:ipd}
I_{\rm pd} = \int \frac{r}{\cst}G\Mp\Sigma K_{3/2}^{(1)}\left|E\right|^2 2\pi r \, \id r,
\end{equation}
and
\begin{equation}
\label{eq:ina}
I_{\rm na} = \int \frac{\alpha}{2} \frac{\Sigma r^2}{\cst} \dd{\cst}{r}e^2 \dd{\varpi}{r}  2\pi r \, \id r
\end{equation}
are two terms related to pressure, one to secular planet-disc interactions and one to non-adiabatic effects, respectively. The coefficient $K_{3/2}^{(1)}$ is defined in Eq. (\ref{eq:ksm}).

For the growth rate we have: 
\begin{equation}
-\Im({\omega})=\frac{S_{\rm ELR} + S_{\rm ECR} + J_{\rm visc} + J_{\rm BC} + J_{\rm na}}{2B}.
\end{equation}
where
\begin{align}
\label{eq:selr}
S_{\rm ELR} & = \sum_{\rm ELR}\int \frac{G\Mp^2}{M_*}\frac{\Sigma}{\cst} |\mathscr{A}E|^2\nonumber\\
&\times w_{\rm L}^{-1}\Delta\left( \frac{r-\rres}{w_{\rm L}} \pm 1 \right) 2\pi r \, \id r,
\end{align}
\begin{align}
\label{eq:secr}
S_{\rm ECR} & = \sum_{\rm ECR}\int\pm \dd{\ln (\Sigma/\Omega)}{\ln r}\frac{G\Mp^2}{M_*}  \frac{\Sigma}{\cst} |\mathscr{C}E|^2\nonumber\\
&\times w_{\rm C}^{-1}\Delta\left( \frac{r-\rres}{w_{\rm C}}\right) 2\pi r \,\id r,
\end{align}
\begin{equation}
\label{eq:jvisc}
J_{\rm visc} = -\int \frac{1}{2}\alpha \Sigma r^2\left|\pd{E}{r}\right|^2 2\pi r \, \id r,
\end{equation}
\begin{equation}
\label{eq:jbc}
J_{\rm BC,i,o} = -\int \frac{\Sigma r^2 \Omega}{\cst} \left|E\right|^2 \frac{R(r)_{i,o}}{\tau_{i,o}} 2\pi r \, \id r,
\end{equation}
and
\begin{equation}
\label{eq:jna}
J_{\rm na} = \int \frac{\alpha}{2}\frac{\Sigma r^2}{\cst} \dd{\cst}{r}r^2 e \pd{e}{r}\,2\pi r \,\id r,
\end{equation}
are terms corresponding to eccentric Lindblad resonances, eccentric corotation resonances, viscosity, boundary conditions and non-adiabatic effects, respectively. Note that for both the precession and growth rates, the non-adiabatic effect can usually be neglected. These relations are slightly different from the ones presented in the main text of \citet{to16} but the steps in their derivation are the same. The two formulations are strictly equivalent but the one presented here is more adapted to locally isothermal discs.

\subsection{Pertubing potential from the planet}
\label{app:phip}
In \citet{to16}, we wrote down the orbit-averaged gravitational potential of a planet of mass $\Mp$ on a circular orbit at radius $\ap$ to be:
\begin{equation}
\Phip = -\frac{G\Mp}{2\ap}\lc{1/2}{0}(\beta),
\end{equation}
where $\beta=r/\ap$, and $\lc{s}{m}$ are the usual Laplace coefficients \citep[see, e.g.,][]{md99}. Using relations between Laplace coeffecients and their derivatives, we find 
\begin{equation}
\dd{\Phip}{r} = -\frac{G\Mp}{2\ap^2}\left[\lc{3/2}{1}-\beta\lc{3/2}{0} \right],
\end{equation}
and
\begin{equation}
\ddd{\Phip}{r} = -\frac{G\Mp}{2\ap^3}\left[ \frac{3}{2}\left(\lc{5/2}{0}+\lc{5/2}{2} \right) - 3\beta\left(2\lc{5/2}{1}-\beta\lc{5/2}{0} \right)  - \lc{3/2}{0}\right].
\end{equation}

Laplace coefficient diverge when $r\to\ap$. It is essential to regularize them to avoid this divergence. Generalizing the work of \citet{to16}, we introduce the softened symmetric kernel $\kc{s}{m}$:
\begin{equation}
\label{eq:ksm}
\kc{s}{m}(r,r') = \frac{rr'}{4\pi} \int_{0}^{2\pi} \frac{\cos m\theta}{\left(r^2 + r'^2 - 2rr' \cos\theta + s^2rr' \right)^s}\id\theta.
\end{equation}

It can be shown that these kernels are related to Laplace coefficients by
\begin{equation}
\kc{s}{m} = \frac{\beta^s}{4(rr')^{s-1}}\lc{s}{m},
\end{equation}
with $\beta$ solution of 
\begin{equation}
\frac{\beta^2+1}{\beta} = \frac{r^2+r'^2}{rr'} + s^2.
\end{equation}
Here $s(rr')^{1/2}$ is the smoothing length, and we take $s=H/r$, the constant disc aspect ratio.

\end{document}